\documentclass[amsmat,amssymb,amsfonts,aps,prb,twocolumn,showpacs,superscriptaddress]{revtex4}

\usepackage{graphicx}
\usepackage{dcolumn}
\usepackage{bm}
\usepackage{color}

\begin{document}

\title{Single exciton spectroscopy of single-Mn doped InAs quantum dots}

\author{J. van Bree}
\affiliation{Departamento de F\'{\i}sica Aplicada, Universidad de Alicante, San Vicente del Raspeig, 03690 Spain}
\affiliation{Department of Applied Physics, Eindhoven University of Technology, P.O. Box 513, NL-5600MB Eindhoven, The Netherlands}
\author{P.M. Koenraad}
\affiliation{Department of Applied Physics, Eindhoven University of Technology, P.O. Box 513, NL-5600MB Eindhoven, The Netherlands}
\author{J. Fern\'andez-Rossier}
\affiliation{Departamento de F\'{\i}sica Aplicada, Universidad de Alicante, San Vicente del Raspeig, 03690 Spain}

\date{\today}

\begin{abstract}

The optical spectroscopy of a single InAs quantum dot doped with a single Mn atom is studied using a model Hamiltonian that includes the exchange interactions between the spins of the quantum dot electron-hole pair, the Mn atom and the acceptor hole. Our model permits to link the photoluminescence spectra to the Mn spin states after photon emission. We focus on the relation between the charge state of the Mn, $A^0$ or $A^-$, and the different spectra which result through either band-to-band or band-to-acceptor transitions. We consider both neutral and negatively charged dots. Our model is able to account for recent experimental results on single Mn doped InAs PL spectra and can be used to account for future experiments in GaAs quantum dots. Similarities and differences with the case of single Mn doped CdTe quantum dots are discussed.

\end{abstract}

\maketitle

%%%%%%%%%%%%%%%%%%%%%%%%%%%%%%%%%%%%%%%%%%%%%%%%%%%%%%%%%%%%%%%%%%
\section{Introduction}

Probing a single magnetic atom in a solid state environement is now possible by
scanning tunneling microscopy (STM), both in metallic \cite{Heinrich07} and
semiconducting surfaces \cite{Paul04,Kitchen06,Paul07,Wiesendanger07}, and by
single exciton spectroscopy in semiconductor quantum dots
\cite{Besombes04,Leger06,Voisin07}, among other techniques. These experiments
permit to address a single quantum object, the spin of the magnetic atom, and to
study its exchange interactions with surrounding carriers.  
Quantum dots doped with a single magnetic atom
are a model system for nanospintronics
\cite{Efros,Govorov05,JFR07}. These systems can also shed light on
the issue of whether or not it is possible to dope
semiconductor nanocrystals  \cite{Nature2005}.

The focus of this work is the single exciton spectroscopy of a single Mn atom in
a InAs quantum dot (QD), motivated by recent experimental results on  InAs QD
\cite{Voisin07} and keeping in mind the relation to previous experiments on
single Mn doped CdTe\cite{Besombes04}. The photoluminescence (PL) spectra of a
CdTe QD doped with only one Mn atom display six narrow peaks, each of which
correspond\cite{Besombes04,JFR06} to one of the six quantum states of the
$S=5/2$ multiplet formed by the 5 Mn $d$ electrons \cite{Furdyna} in Mn$^{2+}$.
Hence, the spin state of the single Mn atom after its interaction with an
exciton in a CdTe QD can be read from the energy and polarization of the emitted
photon \cite{Besombes04,JFR06}. Another interesting result in CdTe dots doped
with Mn, is the fact that the PL spectrum changes radically when a single
carrier is electrically injected into the dot \cite{Leger06}. This has been
explained in terms of the  different effective spin Hamiltonian for the Mn as a
single additional carrier is added into the dot \cite{JFR04,Leger06,JFR07}.

Since Mn is an acceptor in InAs quantum dot, the single Mn exciton PL is expected to be different from CdTe,
where Mn acts as a isoelectronic impurity. Even before photoexcitation, charge neutrality implies that the Mn acceptor in InAs binds a hole. In this neutral acceptor complex, $A^0$, the spin of the Mn is antiferromagnetically coupled to the spin of the acceptor hole, so that $A^0$ behaves as effective spin $F=1$ object. When a neutral exciton $X^0$ is created in a InAs dot doped with 1 Mn, there are 4 spins interacting: the QD electron, the QD hole, the Mn and the acceptor hole. Indeed, recent experimental observations report a band-to-band transition ($X^0A^0 \rightarrow A^0$) PL spectrum with 5 peaks with different intensities at zero applied field, instead of the 6 almost identical peaks in CdTe. The presence of the acceptor hole also opens a new optical recombination channel, the band-to-acceptor transition ($X^0A^0 \rightarrow h^+A^-$) such that the conduction band electron ionizes the Mn acceptor without filling the quantum dot hole.

\begin{figure}[h]
\includegraphics[width=0.8\columnwidth]{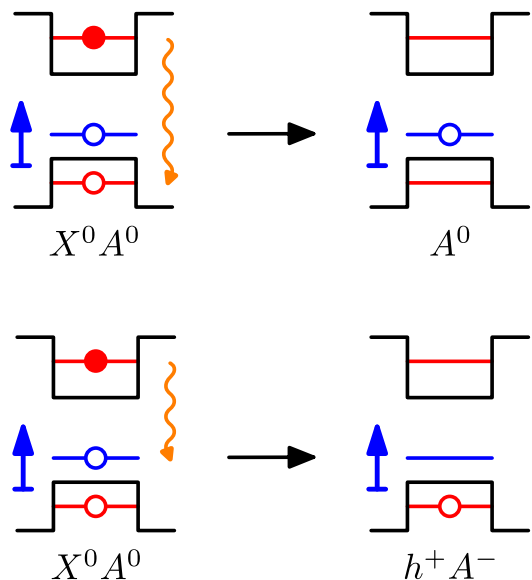}
\caption{Schematic energy levels of the InAs dot doped with 1 Mn. Upper row: band-to-band transition. Lower row: band-to-acceptor transition. (Color online)}
\label{SCHEME1}
\end{figure}

The goal of this paper is to provide a theoretical framework to understand the relation between PL spectra of a single InAs quantum dot doped with one Mn and the interactions, charge and spin state of the relevant degrees of freedom. The rest
of this paper is organized as follows. In section II we describe the theoretical
framework, including the spin models for the relevant degrees of freedom, and
the framework to calculate PL. We discuss both a 4-spin-model and a simpler
2-spin-model, proposed in reference \onlinecite{Voisin07}, and how they are
related. In section III we present our simulations for the PL of neutral quantum
dots. We consider both band-to-band transitions, such that the final state is
the neutral acceptor $A^0$, and the band-to-acceptor transition, such that the
final state is $h^+A^-$, i.e. a QD hole interacting with the spin
$S=5/2$ of the ionized Mn acceptor. We find that an antiferromagnetic QD
hole-Mn coupling can still yield an effective ferromagnetic coupling between the
QD hole and the Mn-acceptor complex (consisting on the Mn ion and the acceptor
hole), as observed by Kudelski {\em et al.} \cite{Voisin07}. In section IV we
present our results for negatively charged quantum dots. In this case, since
there are two electrons and two holes, there are 2 recombination pathways and 4
possible sets of PL spectra are possible. In section V we summarize the main
results.

%%%%%%%%%%%%%%%%%%%%%%%%%%%%%%%%%%%%%%%%%%%%%%%%%%%%%%%%%%%%%%%%%%
\section{Theoretical framework}

\subsection{Photoluminescence and eigenstates}
Our goal is to extract information of the quantum state of the single Mn spin in
the quantum dot from  optical spectroscopy data. We adopt a phenomenological
approach where only the spin degrees of freedom of the Mn, quantum dot (QD)
carriers and acceptor hole are considered. The various spin couplings are chosen
to respect the symmetries of the problem and are fitted to experimental data
when available. The relevant electronic states of the quantum dot are described
in terms of few-spin quantum states which depend on the different spin exchange
interactions.

In the calculation of the PL, it is convenient to distinguish between the ground
state and the exciton state manifolds \cite{JFR06}. Both the number of states
and the effective spin Hamiltonian of these manifolds depend on the charge of
the dot \cite{Leger06,JFR07} and on the charge state of the acceptor complex.
For instance, the relevant degrees of freedom of the ground state manifold (GSM)
of a neutral QD are the spins of the Mn spin and the acceptor hole. The exciton
state manifold (XSM) enlarges the GSM with the addition of the QD electron and
the QD hole. By definition, the GSM is defined by the eigenstates of the
Hamiltonian of the dot before the photoexciton is injected:
\begin{equation}
{\cal H}_G|\Psi_G\rangle=E_G|\Psi_G\rangle
\end{equation}
In the case of Mn in a neutral InAs QD, $G$ runs over the 24 possible states
that can be formed with a spin $S=5/2$ of the Mn and a pseudospin $3/2$ of the
acceptor. Notice that $E_G$ takes different values (ground state spin
splittings) due to either exchange coupling between the Mn and the acceptor hole
or, in the case of compensated impurities or II-VI semiconductors, coupling to an external field. The XSM is formed by the eigenstates of the
exciton Hamiltonian, which can be written as the sum of ${\cal H}_G$ and the new
terms involving all the couplings of the photocarriers with the degrees of
freedom before excitation:
\begin{equation}
{\cal H}_X|\Psi_X\rangle=E_X|\Psi_X\rangle
\end{equation}
Both $E_X$ and $E_G$ and their wavefuntions are obtained from diagonalization of
the model Hamiltonians described below in detail.

The PL spectrum for a given polarization state $\lambda$ is related to these
states\cite{JFR06}:
\begin{equation}
I_{\lambda}(\omega)=\sum_{X,G} n_X %\sum_G
 |\langle \Psi_G|{\cal P}_{\lambda}
|\Psi_X\rangle|^2 \delta\left[\hbar \omega- (E_X-E_G)\right]
\label{PL}
\end{equation}
Here $n_X$ is the probability that a given XSM state is occupied and $\langle
\Psi_G|{\cal P}_{\lambda}|\Psi_X\rangle$ are the the matrix elements, of the interband
electric dipole operator \cite{JFR06} that promotes an electron from the valence
states to the conduction states and viceversa. This operator obeys the standard
optical selection rules associated to the photon with polarization $\lambda$ and
does not affect the Mn spin state. Explicit expressions of this operator are
provided once we discuss the nature of ${\cal H}_G$ and ${\cal H}_X$ and their
eigenvectors.

In a non-magnetic dot, the PL spectrum has a single line at zero magnetic field.
A distinctive feature of magnetically doped dots is the appearance of several
lines at zero magnetic field \cite{Besombes04,Leger06,Voisin07}. According to
equation (\ref{PL}) the appearance of several lines in the PL spectra can occur
both due to splittings in the GSM and in the XSM. The intensity of the lines
depends on two factors, the quantum mechanical matrix elements and the
statistical occupation of the emitting state, $n_X$. This quantity depends on
the complicated non-equilibrium kinetics of the photoinjected carriers. Instead
of solving a non-equilibrium master equation \cite{JFR07}, we assume that the
emitting states are in a thermal equilibrium with an effective temperature which
can be larger than the temperature of the lattice. This phenomenological
approach is supported by experimental results in the case of single Mn in a CdTe
QD \cite{Besombes04}.

\subsection{4-spin-model}

\subsubsection{Ground State Manifold}
Mn has two $s$ electrons which participate in the $sp$-bonding, whereas In has
three. Therefore, substitutional Mn in InAs behaves like an  acceptor. EPR
\cite{EPR01} and photoemission \cite{PES02} experiments indicate that Mn retains
the five $d$ electrons when doping concentrations are small. Hence, Mn keeps an
oxidation state of $+2$ resulting in an effective charge of $-1$, that repels
the electrons nearby. The Mn impurity remains charge neutral, at the scale of a
few unit cells, by binding a hole. The binding energy of the hole is of $110$
meV in Ga(Mn)As \cite{MnGaAs1,MnGaAs2} and $28$ meV in In(Mn)As
\cite{MnInAs,Wiesendanger07}. The acceptor hole state has a radius of approximately 1 nm and
has been probed by STM experiments both in GaAs \cite{Paul04} and InAs
\cite{Wiesendanger07}. In bulk the acceptor hole has a fourfold degeneracy
inherited from the top of the valence band, which is lifted by quantum
confinement and/or strain. Because of the strong spin-orbit interaction, it is
convenient to treat the acceptor hole as a spin $j=3/2$ object, exchange coupled
to the Mn spin $M=5/2$. 

 The operators acting upon this object are the four by four $J=3/2$ angular
momentum matrices, $\vec{j}$. In the spherical approximation \cite{Lipari} the
Mn-acceptor hole spin coupling reads \cite{Benoit}:
\begin{eqnarray}
{\cal H}_{M,j}= \epsilon \vec{M}\cdot\vec{j} \label{benoit}
\end{eqnarray}
where $\epsilon=+5$ meV is the antiferromagnetic coupling between the acceptor hole and
the Mn, $\vec{M}$ are the $S=5/2$ spin matrices of the Mn and $\vec{j}$ are the
$J=3/2$ matrices corresponding to the total angular momentum of the valence band
states.
The Hamiltonian equation (\ref{benoit}) is readily diagonalized in the basis of
the total spin $F=M+J$, the spin of the Mn plus acceptor hole complex. $F$ can take
integer values between 1 and 4. The eigenvalues are $E(F)=\frac{\epsilon}{2}F(F+1) + E_0$. Since the coupling is antiferromagnetic, the ground state has
$F=1$, separated from the $F=2$ states by a relatively large energy barrier of
$2\epsilon$. Hence, as long as the Mn-acceptor hole complex is not distorted by
perturbations that couple different $F$ manifolds and temperature is low enough,
it is a good approximation to think of it as being a composite object with total
spin $F=1$. Since they play an important role, we explicitly write down the 3
wave functions of the $F=1$ manifold:
\begin{eqnarray}
|1,+1\rangle= 
\frac{1}{\sqrt{2}}\left|\frac{5}{2},\frac{-3}{2}\rangle \right. - 
\frac{\sqrt{30}}{10}\left|\frac{3}{2},\frac{-1}{2}\rangle \right.+ \nonumber \\
+\frac{\sqrt{15}}{10}\left|\frac{1}{2},\frac{1}{2}\rangle \right. -
\frac{\sqrt{5}}{10}\left|\frac{-1}{2},\frac{3}{2}\rangle \right.
\nonumber \\
|1,0\rangle= \frac{1}{\sqrt{5}}\left|\frac{3}{2},\frac{-3}{2}\rangle \right.- 
\frac{\sqrt{30}}{10}\left|\frac{1}{2},\frac{-1}{2}\rangle \right. + \nonumber \\ 
+\frac{\sqrt{30}}{10}\left|\frac{-1}{2},\frac{-1}{2}\rangle\right. 
-\frac{1}{\sqrt{5}}\left|\frac{-3}{2},\frac{3}{2}\rangle\right.
\nonumber \\
|1,-1\rangle= 
\frac{-1}{\sqrt{2}}\left|\frac{-5}{2},\frac{3}{2}\rangle \right. + 
\frac{\sqrt{30}}{10}\left|\frac{-3}{2},\frac{1}{2}\rangle \right.- \nonumber \\
-\frac{\sqrt{15}}{10}\left|\frac{-1}{2},\frac{-1}{2}\rangle \right. +
\frac{\sqrt{5}}{10}\left|\frac{1}{2},\frac{-3}{2}\rangle \right.
\label{F1}
\end{eqnarray}
where we use the notation 
$$|F=1,F_z=\pm1,0\rangle=\sum_{M_z,j_z} C_{M_z,j_z}(F,F_z)|M_z,j_z\rangle$$
We see that because of the very strong exchange interaction between the acceptor hole and
the Mn spins, their spins are strongly correlated and they are not good quantum
numbers separately.

Following Govorov \cite{Govorov04}, we assume that the QD  is larger than the bulk acceptor state. Thus, the  QD perturbs weakly the   acceptor state. In this approximation we can distinguish between quantum confined or QD states and acceptor states. The former are  extended all over the dot
the latter are tightly bound to the Mn impurity and their  energy lies  in the gap (see figure \ref{SCHEME1}). The opposite scenario, in which the QD size is comparable or smaller than the acceptor state, has been considered by Climente et al. \cite{Climente05}, would yield different results incompatible with the experiments of Kudelski {\em et al.}\cite{Voisin07}, as discussed below. 

The cubic symmetry of the ideal crystal and the presence of quantum confinement
and strain result in additional terms in the Hamiltonian, that need to be summed
to equation (\ref{benoit}). Both quantum dot confinement and strain can result
in a splitting of the light and heavy hole bands, that can be modelled a $-D
j_z^2$ term. This term would be present in thin film layers with strain and
still preserves rotational invariance in the $xy$ plane. The presence of the
quantum dot potential will break this in plane symmetry. To lowest order
\cite{Govorov} this can be modelled by an additional term in the Hamiltonian, $E
(j_x^2-j_y^2) $. Notice that we assume that these perturbations act on the
acceptor state only and not on the Mn $d$ electrons. This is justified since the
hole is spread over tens of unit cells, whereas the Mn $d$ states are confined
within an unit cell. Hence, we take the following model for the ground state
Hamiltonian:
\begin{eqnarray}
{\cal H}_G= \epsilon \vec{M}\cdot\vec{j} -D j_z^2 + E (j_x^2-j_y^2) \label{DE} 
\end{eqnarray}
As we discuss below we have $\epsilon \gg D \gg E$. Within the $F=1$ lowest
energy manifold the $D$ term splits the triplet into a $F_z=\pm 1$ doublet and a
$F_z=0$ singlet. The $E$ term hybridizes the $F_z=\pm 1$ states, resulting in a
small hybridization splitting.

We obtain the the eigenstates of ${\cal H}_G$ by expressing them as linear
combinations of $|M_z,j_z\rangle=|M_z\rangle \otimes|j_z\rangle$:
\begin{equation}
|\Psi_G\rangle= \sum_{M_z,j_z} {\cal C}_{M_z,j_z}^G |M_z,j_z\rangle =\sum_{F,F_z} {\cal
D}_{F,F_z}^G |F,F_z\rangle \label{GSM-state}
\end{equation}
and diagonalizing numerically the Hamiltonian matrix. The use of the
$|F,F_z\rangle$ basis might be better for interpretation of the results.

Finally, in some instances we need to consider the ionized acceptor complex,
$h^+A^-$. This is the case if we consider the band-to-acceptor transition in
neutral dots or if we consider a charged quantum dot. The spin of the Mn inside
the  $A^-$  state  is  $M=5/2$ and should have properties similar to those of Mn
in CdTe \cite{Besombes04}.

\subsubsection{Exciton state manifold}
We now consider states with an electron and a hole in the QD lowest energy
 levels in the conduction and valence band respectively (see figure
 \ref{SCHEME1}). In contrast to the case of neutral Mn in II-VI semiconductor, the exciton
 states involves 4 spins instead of three: the Mn ($M=5/2$), the QD conduction
 electron $\sigma_c=\pm 1/2$, the QD hole $\sigma_h= \pm 3/2$ and the acceptor
 hole $J=3/2$. Since we ignore LH-HH mixing for the QD valence states, the QD
 holes  are heavy hole, with well defined  $J_z=\pm 3/2$ (or
 $\Uparrow,\Downarrow$).  As a result, the spin couplings of the QD to the other spins (Mn, acceptor hole and QD electron) are Ising like. 
Including the small LH-HH mixing present in the QD hole state   results
in a  small spin-flip terms in the exchange Hamiltonian of the QD hole \cite{JFR07}. 
We label the hole spin states as  the time reversed states of the valence
electronic Bloch states with quantum number $\sigma_v$ \cite{Sham93}, $\sigma_h=-\sigma_v$. With this notation, the spin of a given state that features  one quasiparticle in the valence band, either one electron or one hole, is the same as the spin of the quasiparticle. With this notation, the exciton spin $X$ satisfies the rule $X=\sigma_h+\sigma_c$ and takes values $\pm 1$ states for optically active excitons and $\pm 2$ for optically dark excitons\cite{Sham93,Bayer02}.
%\begin{eqnarray}
%|+2\rangle &=& |\uparrow\rangle_c |\Uparrow\rangle_h \nonumber \\ 
%|+1\rangle &=& |\downarrow\rangle_c |\Uparrow\rangle_h \nonumber \\ 
%|-1\rangle &=& |\uparrow\rangle_c |\Downarrow\rangle_h \nonumber \\ 
%|-2\rangle &=& |\downarrow\rangle_c |\Downarrow\rangle_h 
%\end{eqnarray} 

Since we have 4 spin degrees of freedom we need to consider the 6 two spin couplings between them: 
\begin{eqnarray}
{\cal H}_X={\cal H}_G + {\cal H}_{c,h} + {\cal H}_{h,M} +{\cal H}_{h,j} + {\cal H}_{c,M} 
+{\cal H}_{c,j} + {\cal H}_{Z} \label{HAMIL}
\end{eqnarray}
One of them, the $\vec{M}\cdot\vec{j}$ term, is present both in the GSM and XSM.
The symmetry and the coupling strength characterize a given spin-spin
interaction. 
In spin rotational invariant systems two spins $\vec{s}_1$ and
$\vec{s}_2$ interact via Heisenberg coupling, $\vec{s}_1\cdot\vec{s}_2$.
 When the interplay of 
spin-orbit coupling and lack of spherical symmetry break spin rotational 
symmetry, spins are coupled with
different strengths along different directions.
An extreme case are flat self-assembled quantum dots, for which 
the lowest energy hole states are purely heavy holes such that
in-plane couplings are strictly
forbidden\cite{JFR06}, resulting in Ising couplings. 
In the opposite limit, the
conduction band states and the Mn $d$ states have no orbital momentum which
greatly reduces the size of spin-orbit interactions, resulting in Heisenberg
couplings between each other. 
An intermediate situation would be that of holes in spherical nanocrystals,
where, in spite of strong spin orbit interactions, 
Mn-hole exchange is still described with a Heisenberg coupling 
 \cite{Bhatta03}. 
Following previous work in CdTe
\cite{JFR06,JFR07}, we take the QD hole-Mn and the QD electron-Mn coupling as
antiferromagnetic Ising and ferromagnetic Heisenberg respectively.

The second term in equation (\ref{HAMIL}) is the longitudinal QD electron-hole exchange that splits
the bright $\pm 1$ and dark $\pm 2$ excitons in two doublets:
\begin{equation}
H_{eh}= +J_{eh}\sigma_c \sigma_h
\end{equation}
Since the dark doublet, for which $\sigma_c\sigma_h>0$  
is lower in energy we have $J_{eh}<0$.   For
simplicity, we neglect transverse electron-hole exchange.

In the case of $A^0$ we also need to include the coupling of the QD electron and QD hole spins both to
the Mn spin and to the acceptor hole. We assume the same symmetry for the two couplings:
\begin{equation}
{\cal H}_{hM}+{\cal H}_{hj}= J_{hM}\sigma_h M_z + J_{hj}\sigma_h j_z 
\end{equation}
Notice  that
the sign of the hole-hole coupling is not clear a priori. 
The QD conduction electron couplings are:
\begin{equation}
{\cal H}_{cM}+{\cal H}_{cj}= -J_{cM}\vec{\sigma}_c \cdot\vec{M} - 
J_{cj}\vec{\sigma}_c \cdot\vec{j}
\end{equation}

Finally, we include  the Zeeman coupling of the various spins to an external magnetic
field. For simplicity, we ignore the orbital coupling to the magnetic field. In the neutral dot
there are 2 spins in the GSM and 4 spins in the XSM that are coupled to the magnetic field. Thus,
we need 4 $g$-matrices. In this paper we only consider magnetic fields along the growth axis. The
couplings read:
\begin{eqnarray}
{\cal H}_Z= \mu_B B_z \left(g_c\sigma_c + g_h \sigma_h + g_M M_z + g_j j_z \right) \label{Zeeman}
\end{eqnarray}
where $\mu_B= \frac{\hbar e}{2 m} = +0.0579$ meV/T.

The exciton states in the 4-spin-model are obtained by numerical diagonalization of the
Hamiltonian. We express them as linear combinations of the product basis
$|M_z,j_z,\sigma_c,\sigma_h\rangle$:
\begin{equation}
|\Psi_X\rangle= \sum_{M_z,j_z,\sigma_c,\sigma_h} 
{\cal C}_{M_z,j_z,\sigma_c,\sigma_h}^X |M_z,j_z,\sigma_c,\sigma_h\rangle 
\end{equation}
The 4-spin-model has 96 eigenstates, as many as the product of the $(2S+1)\times(2J+1)\times4$,
where $2S+1=6$ is the multiplicity of the Mn spin, $2J+1=4$ is the multiplicity of the acceptor
hole spin, and 4 is the number possible of quantum dot exciton states.

The eigenvalues and eigenvectors of ${\cal H}_X$ depend on the strength of the spin couplings.
These depend both on the material and the sample. For instance, exchange coupling between the QD
carriers to the Mn spin is given by the product of the material exchange integrals\cite{Furdyna},
$\alpha$ and $\beta$, and the probability amplitude of the QD envelope functions for either
electrons or holes \cite{JFR04,JFR06}, which is clearly a sample dependent property. For the same
reason, the hole-Mn coupling is much stronger for the acceptor state than for the QD state. Here we
choose the numerical values of the exchange coupling constants to account for the experimental
data. Sample to sample variations will result in different PL spectra. 

Importantly, as long as the Mn-acceptor hole is the dominant coupling, there are 12 lowest energy
exciton states, well separated from the rest. These 12 states correspond to the possible
combinations of the 4 exciton states and the 3 $F=1$ states. Although these states are
predominantly $F=1$, they are somewhat mixed with higher $F$ states. It must be noted that,
assuming a thermal occupation with an effective temperature, the PL is predominantly given by
transitions from the $F=1$ manifold.

% The occupation of the higher energy states 
% with $F>1$ within the XSM is presumably smaller and they give a small
% contribution to the PL.

\subsection{2-spin-model}
The 4-spin-model has as much as 6 exchange constants which might not be possible to extract from
comparison with PL experiments. The situation can be significantly simplified trading off some
accuracy. In the GSM, we can remove the $F>1$ states as long as they are not thermally occupied
($k_B T < 2 \epsilon$) and not mixed dynamically through the terms that break rotational
symmetry, i.e. as long as both $D$ and $E$ are also much smaller than $2\epsilon$. If these
conditions are met, we can use a single spin model for the ground state \cite{Voisin07}. The 3 by 3
Hamiltonian in the $F=1$ subspace reads: 
%\begin{eqnarray}
%H_G = \left(\begin{array}{ccc} -D & 0 & E \\ 0 & 0 & 0\\ E & 0 &-D\end{array}
%\right) \end{eqnarray}
\begin{eqnarray}
{\cal H}_G= -{\cal D} F_z^2 + {\cal E} (F_x^2-F_y^2) + g_F\mu_B F_z B_z
\end{eqnarray}
A Hamiltonian similar to this has been used to model Mn in GaAs quantum
wells\cite{Sapega07}.
The coupling constants of the   2-spin-model are obtained from those of the 4-spin-model by
 representing equations (\ref{DE}) and (\ref{Zeeman}) in the basis
set of the $F=1$ states from equation (\ref{F1}). We obtain
\begin{eqnarray}
{\cal D}&=&\frac{3}{10}D \nonumber \\
{\cal E}&=&\frac{3}{10}E\nonumber \\
g_F&=&\frac{7}{4} g_M - \frac{3}{4} g_j
\label{from2to4}
\end{eqnarray}

The ${\cal D}$ term splits the $F=1$ triplet into
a singlet $|1,0\rangle$ and a doublet, $|1,\pm1 \rangle$. The ${\cal E}$ term mixes the two states
in the $\pm 1$ doublet resulting in a bonding and anti-bonding states along the $Y$ and $X$ axis respectively.

Along the same lines, the XSM Hamiltonian can be approximated by a simpler 2-spin-model one, if we treat
the optically active exciton as a quantum Ising degree of freedom, $X_z=\pm 1$, coupled to the spin
$F=1$ formed by the Mn spin and the acceptor hole. In that case, the XSM Hamiltonian reads:
\begin{equation}
{\cal H}_X = {\cal H}_G + {\cal J} F_z X_z + g_X \mu_B X_z B_z \label{2spinX}
\end{equation} 
The strength of the effective exciton Mn complex coupling is related to the bare coupling constants
of the 4-spin-model through:
\begin{eqnarray}
{\cal J} &=& \frac{21}{8} J_{hM} - \frac{9}{8} J_{hj} \nonumber \\
g_X &=& \frac{3}{2} g_{h} - \frac{1}{2} g_{c} \label{sign-reverse}
%\quad \text{or} \quad g_T = \frac{3}{2} 
\end{eqnarray}
In obtaining equation (\ref{sign-reverse}) we set the electron-Mn complex interaction to zero. It is worth noting
that the sign of $g_F$, the effective $g$-factor of the effective composite spin $F$, as well as
the sign of ${\cal J}$, the effective exciton-Mn complex coupling, could be different from those of
the constituent particles. In particular, even if both the QD hole-acceptor hole $J_{hj}$ and the
QD hole-Mn couplings $J_{hM}$ are antiferromagnetic, we could have a negative (ferromagnetic)
effective coupling if $J_{hj}>\frac{7}{3}J_{hM}$. This sign reversal occurs because the very strong
Mn-acceptor hole interaction distorts their wave functions and perturbs their couplings to a third
spin.

\begin{figure}[hbt]
\includegraphics[width=\columnwidth]{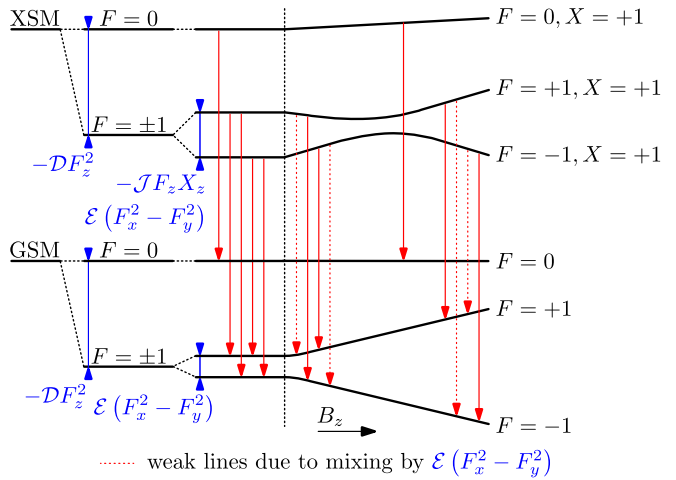}
\caption{Energy level diagram for the neutral exciton band-to-band transitions. Both the ground
state and exiton state ($X=+1$) energy lines are shown, as well as their evolution as a function of the
magnetic field. (Color online)}
\label{X0A0_A0-diag}
\end{figure}

The advantage of the 2-spin-model is that it can be solved analytically. The details are provided
in the appendix. Importantly, the $F_z=0$ state is decoupled from the $F_z=\pm 1$ pair, both in the
GSM and the XSM. The energy level diagram is shown in the figure (\ref{X0A0_A0-diag}). The GSM
features a weakly split doublet. The energy separation with the higher energy $F_z=0$ singlet is
${\cal D}$. The doublet, denoted $x$ and $y$, is a linear combination of the $F_z=\pm 1$ states.
The small splitting within the doublet is aproximately equal to ${\cal E}$. The Zeeman term affects
the $F_z=\pm 1$ states.

Since in the 2-spin-model the $F_z=0$ state is decoupled from the $F_z=\pm 1$ states,  it is
convenient to think of the doublet $|F=1,F_z=\pm 1\rangle$ as a isospin $1/2$ space. Both the
exchange coupling and the applied field (in the Faraday geometry) act as effective magnetic fields
along the isospin $z$ axis, whereas the ${\cal E}$ term acts an effective magnetic field in the
isospin $xy$ plane. In the 2-spin-model the bright exciton does not shift the $F_z=0$ state. Thus,
the exciton exchange and the magnetic field mix the $x$ and $y$ wave functions of the $F_z=\pm 1$
doublet. The mixing opens two new optical transitions, marked in the diagram of figure
(\ref{X0A0_A0-diag}).

\subsection{Optical selection rules}
We now discuss the selection rules associated to the dipole ${\cal P}_{\lambda}$ operators that we need to
use in equation (\ref{PL}). They promote an electron from the valence band to the conduction band
and viceversa. In a Mn doped III-V QD there are two relevant valence band levels so that exciton
recombination can occur through two channels, as shown in figure \ref{SCHEME1}: band-to-band and
band-to-acceptor. These transitions have different energies and, since the wavefunctions of these
holes are not the same, different optical selection rules.

Emission of a $\sigma^+$ ($\sigma^-$) photon in the direction normal to the QD layer takes away
(adds) one unit of angular momentum $L_z$ from the system. In the band-to-band transitions we shall
ignore LH-HH mixing of the quantum dot hole state. Therefore, emission of a $\sigma^+$ ($\sigma^-$)
implies the removal of the $+1$ ($-1$) exciton. Hence, in the band-to-band channel, the
transition operators are defined from its action upon XSM states as a projector:
\begin{eqnarray}
{\cal P}_{-}|\Psi_X\rangle &=& \sum_{M_z,j_z} 
{\cal C}^X_{M_z,j_z,\uparrow ,\Downarrow} |M,j_z\rangle
\nonumber \\
{\cal P}_{+}|\Psi_X\rangle &=& \sum_{M_z,j_z} {\cal C}^X_{M_z,j_z,\downarrow ,\Uparrow}
|M_z,j_z\rangle \label{dip1}
\end{eqnarray}
We omit the prefactor proportional to the single particle dipole matrix element, which is a
convolution of the atomic and the envelope wave functions. Ignoring LH-HH mixing in the
band-to-band transition implies that, in our model, linear polarization can only occur through
quantum coherence between the $+1$ and $-1$ exciton.

In the band-to-acceptor transition the dipole operator moves an electron from
the QD conduction level to the acceptor level, for which we can not ignore LH-HH
mixing. Hence the gain (loss) of one unit of angular momentum upon $\sigma^+$
($\sigma^-$) photon emission can occur also through the light hole channel.
Thus, in a band-to-acceptor transition the spin of the annihilated conduction
band electron $\sigma_c$ and the acceptor hole $j_z$ are given to the
$\sigma^{\pm }$ photon:
\begin{equation}
\sigma_c+j_z =\pm 1
\label{rule1}
\end{equation}
The  band-to-acceptor transition operator ${\cal P}_{\pm}$ 
is fully described  by its action upon a given state $|\Psi_X\rangle$
of the XSM. The resulting GSM state read, for $\sigma^+$ emission:
\begin{eqnarray}
{\cal P}_{+}|\Psi_X\rangle = 
\sum_{M_z,\sigma_h} {\cal C}^X_{M_z,+3/2,\downarrow,\sigma_h} 
|M_z\rangle \otimes |\sigma_h\rangle + \nonumber \\ 
+ \frac{1}{\sqrt{3}}{\cal C}^X_{M_z,+1/2,\uparrow,\sigma_h}
 |M_z\rangle \otimes |\sigma_h\rangle
 \label{b2aplus}
\end{eqnarray}
and for $\sigma^-$ emission:
\begin{eqnarray}
{\cal P}_{-}|\Psi_X\rangle = 
\sum_{M_z,\sigma_h} {\cal C}^X_{M_z,-3/2,\uparrow,\sigma_h} 
|M_z\rangle \otimes |\sigma_h\rangle + \nonumber \\ 
+ \frac{1}{\sqrt{3}}{\cal C}^X_{M_z,-1/2,\downarrow,\sigma_h}
 |M_z\rangle \otimes |\sigma_h\rangle
 \label{b2aminus}
\end{eqnarray}

 Notice that these
operators leave the quantum dot hole unchanged and connect states
in which the Mn spin is strongly coupled to the acceptor hole to states where the 
acceptor hole is compensated and the Mn spin is only coupled to the QD hole.
 Notice that, both in band-to-band and band-to-acceptor transitions,
the dipole operators do not act on the Mn $d$ electrons. Therefore, the spin of
the Mn is conserved during the photon emission processes. This is in contrast to
intra-shell transitions relevant when the bandgap is larger than the
intra-atomic transitions \cite{Birman03}.

%%%%%%%%%%%%%%%%%%%%%%%%%%%%%%%%%%%%%%%%%%%%%%%%%%%%%%%%%%%%%%%%%%
\section{Neutral exciton spectroscopy}

We now present our calculations for the PL spectra for neutral InAs quantum dots doped with one Mn. We consider both band-to-band $X^0A^0 \rightarrow A^0$ and band-to-acceptor $X^0A^0 \rightarrow h^+A^-$ transitions. The symmetry of the state left behind after photon emission is very different in these two cases. In reference \onlinecite{Voisin07} the band-to-band transition $X^0A^0 \rightarrow A^0$ has been experimentally observed and described with a model very similar to the one presented in the previous section. We first revisit this case which permits to obtain numerical values for the parameters in the Hamiltonian. This makes it possible to address the band-to-acceptor transition for which there is no experimental data and no available prediction of how the PL spectra should look like.

\subsection{Band-to-band transitions: 4-spin-model}
\begin{figure}[h]
\begin{tabular}{c}
 \includegraphics[width=\columnwidth]{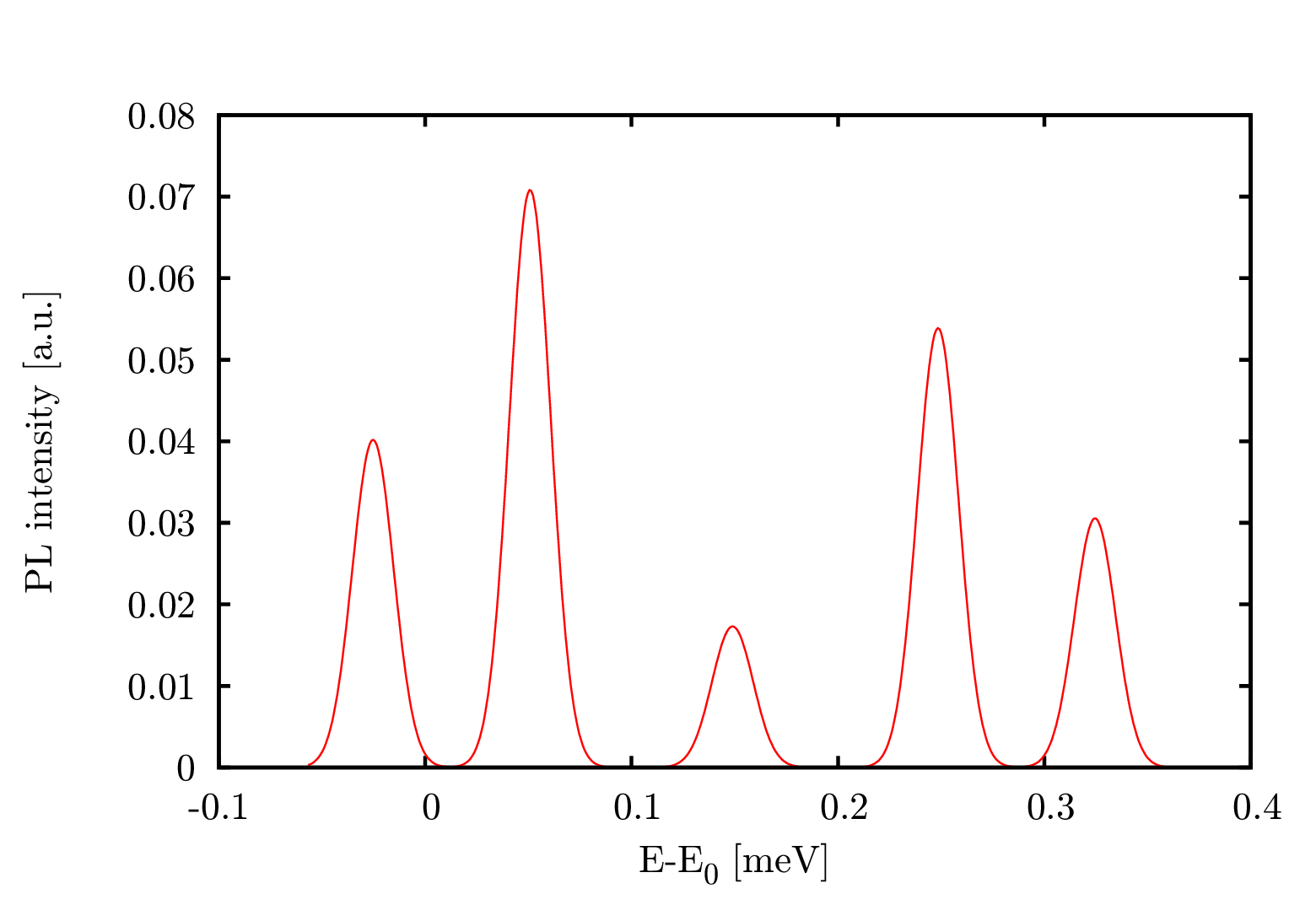} \\ 
 \includegraphics[width=\columnwidth]{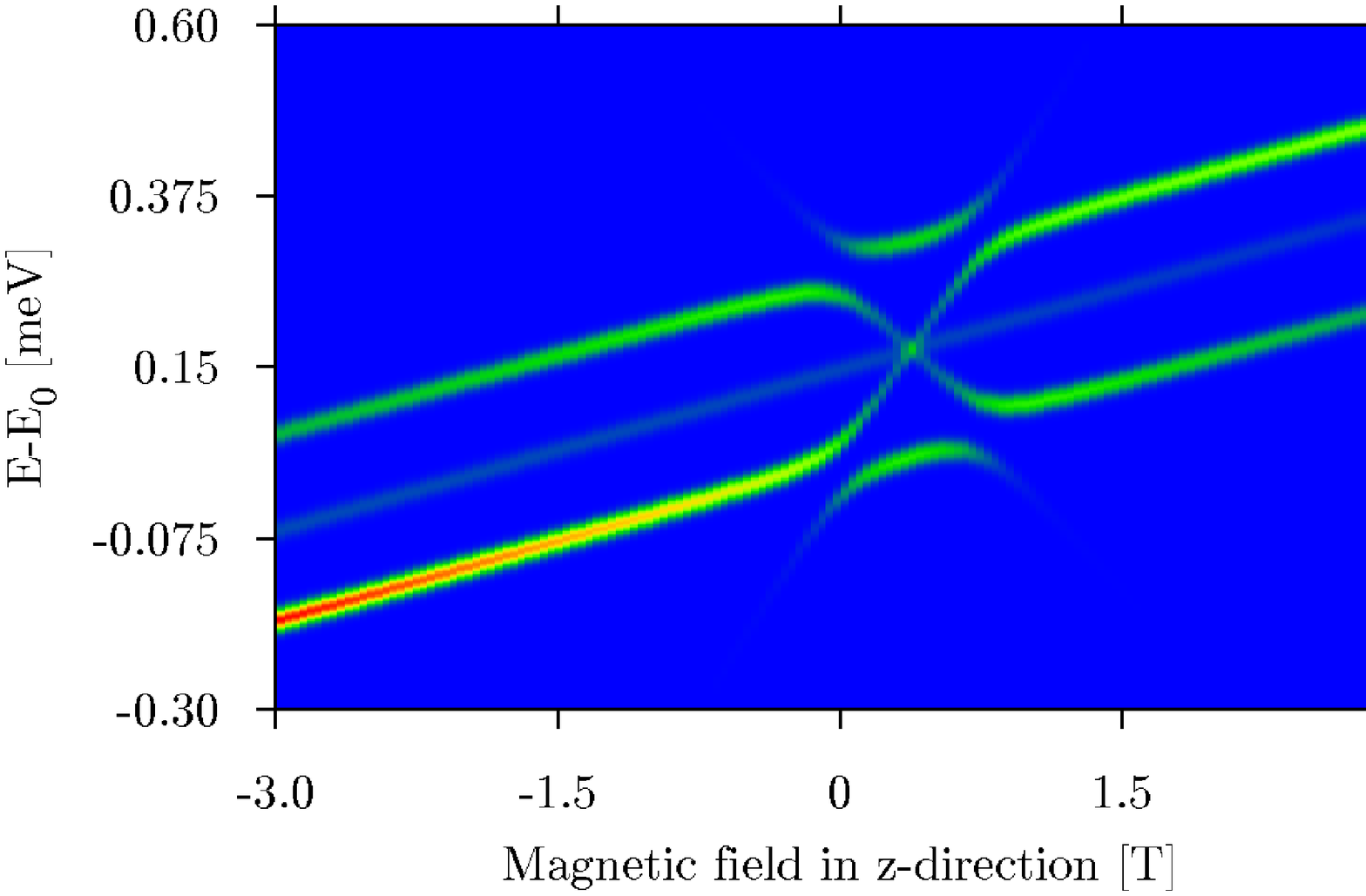} 
\end{tabular}
\caption{Neutral exciton band-to-band transition calculated with the 4-spin-model. Upper panel: $B_z=0$ $\sigma^+$-PL. Lower panel: color plot of $\sigma^+$-PL intensity as function of energy (vertical axis) and applied magnetic field (horizontal axis). (Color online)}
\label{X0A0PL}
\end{figure}

In the band-to-band transition the system is left in one of the GSM states discussed in the previous section where the Mn-acceptor hole complex behaves like a $F=1$ spin. In band-to-band transitions the spin $A^0$ complex is probed by the quantum dot exciton. This is the scenario considered by Kudelski {\em et al.} \cite{Voisin07}. Here we compute the PL spectrum by numerical diagonalization of the GSM and the XSM within the 4-spin-model and then combining equations (\ref{PL}) and the polarization operator equation (\ref{dip1}). 
We use  the numerical values of the coupling constants in the 4-spin-model as to fit the experimental PL spectrum of reference \onlinecite{Voisin07}. The experimental zero field PL features 5 lines: a central low intensity one in between a high and a low energy doublet. Both the $B_z=0$ zero field PL and the PL as a function of energy and applied field in the Faraday geometry are shown in figure \ref{X0A0PL}. Our calculations, shown in figure (\ref{X0A0PL}), reproduce the zero field PL\cite{Voisin07} with 5 peaks at zero magnetic field, as well as the main features of the PL as a function of magnetic field. Notice that in the horizontal axis we plot with respect to $E_0$ the transition energy of the bare quantum dot exciton, excluding its coupling to the Mn.

The origin of the 5 peaks at zero field can be understood by inspection of the
energy diagram shown in figure (\ref{X0A0_A0-diag}) (see also reference
\onlinecite{Voisin07}), in which we only show states within the $F=1$ manifold.
The basic idea is that the quantum dot exciton is probing the $F_z$ component of a 
spin $F=1$ object. If $F_z$ was a good quantum number, 3 lines should be
expected: the middle peak, corresponding to $F_z=0$, and a high and low energy
peaks, corresponding to the spin splitting of the $F_z=\pm 1$ states Ising
coupled to the exciton spin. However, the in-plane anisotropy results in the
mixing states within the $F_z=\pm 1$ doublet into $x$ and $y$ states with
slightly different energies. As a result, there are direct ($xx$,$yy$)
transitions as well as crossed transitions ($xy$,$yx$). The fact that the height
of the $xx$ and $xy$ transitions are similar denotes that the exchange
interaction is comparable to the in-plane anisotropy term $E$. The energy
difference between these satellite ($F_z=\pm 1$) transitions and the $F_z=0$
central peak arises from the exchange coupling to the exciton.

The origin of the reduced intensity of the $F_z=0$ central peak is not in the
quantum mechanical matrix elements, but in the smaller statistical occupation
probability, given that the $F_z=0$ state has a higher energy than the $F_z=\pm 1$ doublet.
Interestingly, since there are no transitions mixing $F_z=\pm 1$ to $F_z=0$, $D$
can not be inferred directly from PL line splittings. In contrast, the value of
$D$ stronlgy affects the intensity of the central peak. We estimate $D\simeq4$
meV.

These results are different from single Mn doped CdTe quantum dots, for which
the PL has 6 peaks at zero field. There the QD exciton is probing the $M_z$
component of a spin $5/2$ object without in-plane magnetic anisotropies. Here
the 5 peaks show the interaction of a QD exciton with a spin $F=1$ object with
in-plane magnetic anisotropy.

Additional information is obtained from the evolution of the PL spectra as a
function of an applied magnetic field along the growth axis $z$. In the
experiment \cite{Voisin07} the 5 lines seen at zero field evolve, changing both
in intensity and energy, in an intricated manner. In figure the lower panel of
(\ref{X0A0PL}) we show a contour plot of the PL intensity as a function of
energy (vertical axis) and magnetic field (horizontal axis) for $\sigma^+$
transitions, obtained within the 4-spin model. The fact that the calculation is
in fairly good agreement with the experiment, and provides a strong back up for
the theory.

\subsection{Band-to-band transitions: 2-spin-model}
We now discuss the physical interpretation of the evolution of the PL spectra as
a function of the applied magnetic field using the 2-spin-model proposed by Kudelsy {\em et al.}\cite{Voisin07}. 
The
2-spin-model affords analytical expressions for the PL spectrum at finite
magnetic field in the Faraday configuration. The derivation is shown in the
appendix.
We address the merger of 3 lines at a particular value of the applied field,
$B^*$, the non-monotonic evolution of the highest and lowest energy lines at
small field and the quenching of their intensity at large fields. 
 Within this model a given circular polarization of the photon fixes
the QD exciton spin. There are 3 exciton states and 3 ground states. The $F_z=0$
state, both for the XSM and the GSM, is decoupled from the other states and
gives rise to the central line. The Zeeman shift of this line is that of the QD
exciton, $g_X \mu_B B_z$. Thus, we can fit the experimental data \cite{Voisin07}
and infer from here the $g$-factor of the QD exciton $g_X=1.2$ not far from
values reported before \cite{nakaoka04}.

The other four lines come from transitions within the $F_z=\pm 1$ doublets. The
energy levels of the ground state $F_z=\pm 1$ doublet are given by $-{\cal D}\pm
h_G$ where 
\begin{equation}
h_G=\sqrt{{\cal E}^2+ (g_F \mu_B B_z)^2}
\end{equation}
The energy levels of the $F_z=\pm 1$ doublet in the XSM with exciton spin $X_z=\pm 1$ are given by $E_0-{\cal D}+ g_X\mu_B X_z B_z \pm h_X$ where $E_0$ is the exciton energy transition without spin and Zeeman terms and
\begin{equation}
h_X=\sqrt{{\cal E}^2+ (g_F \mu_B B_z+{\cal J} X_z)^2}
\end{equation} 
Both in the GSM and XSM the two states in the doublet are linear combination of both $F_z=+ 1$ and $F_z=-1$. The mixing between the $F_z=\pm 1$ states is governed by the competition between the in-plane anisotropy ${\cal E}$, and the longitudinal interactions of the $F_z$ spin with the applied field and, in the exciton manifold, the exchange coupling to the exciton. This competition can be described by two angles, $\cot\theta_G=\frac{g_F\mu_B B_z}{{\cal E}}$ and $\cot\theta_X=\frac{g_F\mu_B B_z + {\cal J} X_z}{{\cal E}}$. Due to the different mixing in the GSM and XSM, 4 transitions are allowed. We label them with $ba$, where $a=\pm$ and $b=\pm$ label the low $(-)$ and high $(+)$ energy state of the ground and exciton state in the $F_z=\pm 1$ manifold. From equation (\ref{2spin-energies}) we have the transition energies
\begin{equation}
E^{b\rightarrow a}= E_0 + g_X \mu_B X_z B_z + b h_X-a h_G
\label{2spin-energies-2} 
\end{equation}

At $B_z=0$ we have $h_X=\sqrt{{\cal E}^2+ ({\cal J} X_z)^2}$ and $h_G={\cal E}$,
so that $h_X-h_G>0$. Thus, from the $B_z=0$ point we immediately can label the 4
non-monotonic lines from low to high energy, at zero field, as $-+$ (1), $--$
(2), $++$ (3) and $+-$ (4). We also denote with (0) the central line with
$F_z=0$. The splitting between the two low (high) energy lines is $2h_G$ and
their average energy is $-h_X$ $(+h_X)$. At zero field $h_G={\cal E}$, thus
${\cal E}$ is half the splitting within both the low and the high energy
doublets. We thus infer ${\cal E}=0.035$ meV. The splitting between the high
energy doublet and the low energy doublet is, at zero field, $2h_X= \sqrt{{\cal
J}^2+{\cal E}^2}$. From here we infer $|{\cal J}|=0.14$ meV.

Equation (\ref{2spin-energies-2}) permits to extract the field $B^*$ at which lines (2), (3) and (0) cross. The crossing arises from the compensation between the Zeeman splitting of the $F$ spin and its exchange coupling to the exciton. The condition is $h_X=h_G$ is satisfied for 
\begin{equation}
2 g_F \mu_B B^* = -{\cal J} X_z
\label{Bstar}
\end{equation}
Since $B^*$ is positive for $X_z=+1$ we immediately see that ${\cal J}$ must be negative: the $F_z$ spin is ferromagnetically coupled to the QD exciton. As discussed earlier, the negative sign can be obtained even if in the 4-spin-model the QD hole-Mn coupling is antiferromagnetic. Thus, the negative sign comes from the strong correlation between the Mn spin and the acceptor hole spin, such that the sign of $F_z$ and $M_z$ are anticorrelated in the $F=1$ manifold. In the simulations with the 4-spin model we have used positive values for the QD hole-Mn coupling, obtaing good agreement with the experiment. 
%Taking $g_F=2.77$ and ${\cal J}=0.14$meV, we have $B^*\simeq0.4$T.
Since we can infer  ${\cal J}$ from the $B_z=0$ data,  equation (\ref{Bstar}) permits to
infer $g_F$ from the experimental value of $B^*$. We obtain $g_F=3.01$, not far
from the $g_F=2.77$ of Mn-acceptor complex in GaAs. 

\begin{figure}[h]
\includegraphics[width=\columnwidth]{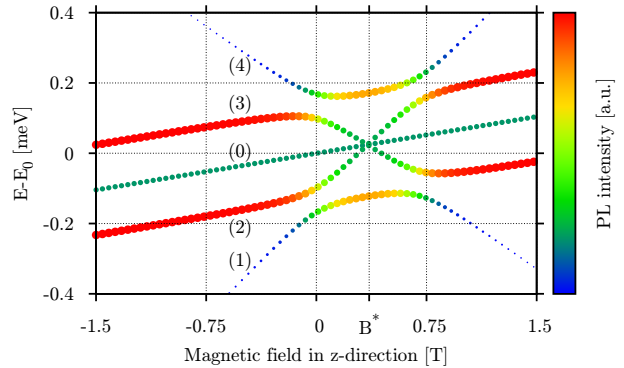}
\caption{$\sigma^+$-PL intensity of the neutral exciton band-to-band
transition, as a function of the magnetic field, calculated with the
2-spin-model (eq. \ref{A12}). Thermal effects are not included.
The size and color of the symbols of the numbered lines is proportional to
the quantum yield. The intensity of the central line is constant. (Color online)
}
\label{b2b-model}
\end{figure}

The intensity of the 4 lines in the $F_z=\pm 1$ manifold are a function of
$\alpha\equiv\frac{1}{2}\left(\theta_G-\theta_X\right)$ (see equation
\ref{table-PL-I}). In particular, the strength of $++$ and $--$ transitions is
given by $\cos^2(\alpha)$ whereas the strength of the $+-$ and $-+$ transitions
is given by $\sin^2(\alpha)$. In the high field limit, when $g_F \mu_B B_z \gg
|{\cal J} X_z|$, we have $\theta_X=\theta_G$ and $\alpha$ goes to zero. In
figure (\ref{b2b-model}) we plot the $\sigma^+$ PL spectra. The size and color of the
circles is proportional to the quantum yield equation (\ref{table-PL-I}). The
quantum yield of the $F_z=0$ transition is constant. The slope
of this line, coming from the Zeeman splitting of the QD exciton, is also
present in the other four lines.

Lines (1) and (4) come from transitions that mix $+$ and $-$ states with
different symmetry. Their energy with respect to line (0) is given by $\pm(h_X+
h_G)$. Since ${\cal J}<0$, $h_X$ has a minumum at $B_z=B^*$ whereas $h_G$, whose
contribution is smaller, has a minimum at $B_z=0$. The intensity of these lines
quenches as the magnetic field increases so much that $g_F \mu_B B_z \gg {\cal E}$
and the $F_z$ is restored as a good quantum number. In contrast, lines (2) and
(3) come from $++$ and $--$ transitions. 
%At small field there is a destructive
%interference of the crossed terms that reduces the quantum yield of these lines
%in favor of the (1) and (4) transitions. 
 At large fields the quantum yield is
increased since the mixing between $F_z=+1$ and $-1$ is quenched. The energy of
(2) and (3) with respect to (0) is given by $\pm(h_X- h_G)$. Thus, when the
Zeeman splitting is much larger than the exchange coupling, i.e. for $B_z \gg
2 B^*$ , we have $h_X\simeq h_G$ and the slope of lines (2) and (3) is the same
as (0).

The model  captures the main experimental features of the PL spectrum \cite{Voisin07}, namely:
{\em i)} 5 peaks distributed 
as a high and low energy doublets with a small intensity central peak;
{\em ii)} as a magnetic is applied along the growth direction, the central line does not change intensity and has a linear shift, whereas the doublets have non-monotonic shifts and do change intensity, two of them fade away;
{\em iii)} at a given value of $B^*$  3 lines, coming from the low and high doublets and the central line, cross.

The results of the 2-spin-model (figure \ref{b2b-model}) and 4-spin-model (figure \ref{X0A0PL}) have no apparent differences (besides the lack of thermal occupation in the 2-spin-model). 
This validates the approximations made to go from the 4-spin-model to the 2-spin-model.
%Upon relaxation of one of these assumptions (e.g. upon applying a very large magnetic field), the 
%$F>1$ states may become important, and the 2-spin-model will fail. 
% Remark: In fact, the quantum number $F$ is actually not a proper quantum number in presence of 
%other couplings to the Mn-acceptor hole complex (albeit that these perturbations are really small
% compared to the coupling strength between Mn and the acceptor hole)

The model portraits the neutral Mn acceptor complex in InAs as a spin $F=1$ nanomagnet with two almost
degenarate ground states, $F_z=\pm 1$, a rather large single ion magnetic
anisotropy of ${\cal D}$, and a small in-plane anisotropy ${\cal E}$. The
quantum dot exciton is Ising coupled to $F_z$ and permits a direct measurement
of ${\cal E}$ and ${\cal J}$, and indirect measurement of ${\cal D}$.  Finally, we have verified that the model proposed by Climente {\em et al.} \cite{Climente05} could {\em not} account for the experimental PL reported by Kudelski {\em et al.}. 
In a nutshell, this model is very close to the one proposed  by one of us to account for the PL of Mn doped charged CdTe quantum dots \cite{Leger06}. In the model of Climente {\em et al.}, the ground state manifold has 6 doublets coming from the Ising coupling of the hole and the Mn spin. The injection of an additional electron-hole pair will result in two holes with opposite spin
occupying the same orbital, uncoupled from the Mn, which would interact only with the electron, presumably via a Heisenberg coupling. Thus, the exciton manifold of such a model would have two spectral lines. The resulting spectrum would have 11 lines with a characteristic V shape 
\cite{Leger06}. If the the Mn-electron coupling is turned off, the model yields 6 equally strong lines.  

\subsection{Band-to-acceptor transition}
We now consider band-to-acceptor transitions, $X^0A^0\rightarrow h^+A^-$ such
that, after photon emission there is a hole left in the QD levels and the Mn is
liberated from the acceptor hole. This kind of transition has been observed in
Mn doped GaAs bulk \cite{kim05} and quantum wells \cite{Awschalom08} but not yet in Mn doped
quantum dots. Three obvious differences with the band-to-band transitions can be
mentioned beforehand. Firstly, the PL spectrum associated to the acceptor
transition should be red-shifted with respect to the band-to-band transition,
by the sum of the acceptor binding energy and the quantum dot confinement
energy. In GaMnAs quantum wells the reported shift is approximately 107 meV
\cite{Awschalom08}. Secondly, we expect a smaller intrinsic efficiency of the
band-to-acceptor process compared to the band-to-band case, due to the smaller
electron-hole overlap in the case of the former. Thirdly, the final state,
$A^-h^+$ is an excited state since the QD hole could be promoted to the acceptor
state, reducing the energy of the system. The spin of the $A^-h^+$ state is the
product of the Mn spin $S=5/2$ and the QD hole spin.

\begin{figure}[hbt]
\includegraphics[width=\columnwidth]{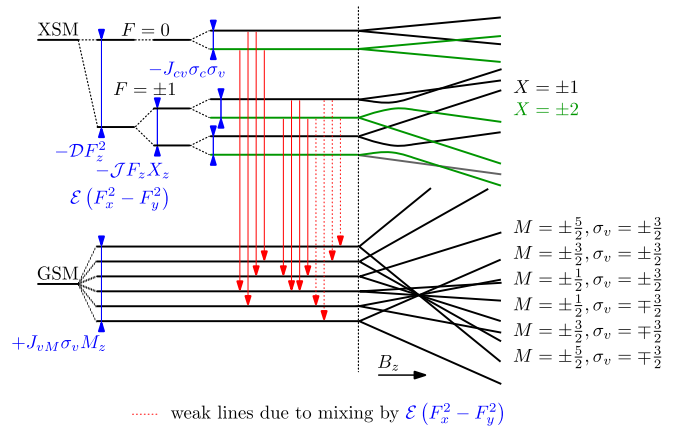}
\caption{Energy level diagram for the neutral exciton band-to-acceptor transitions. Both the ground state and exiton state energy lines are shown, as well as their evolution as a function of the magnetic field. (Color online)}
\label{bandtoacceptor-scheme}
\end{figure}

A diagram of the energy levels of this transition is shown in figure
(\ref{bandtoacceptor-scheme}). The states of the XSM are the same as in the
band-to-band transition, except for the fact that now both the $X=\pm 1$ and the
$X=\pm 2$ transitions are allowed, since the optical selection rule must be
enforced with the acceptor hole and not with the QD hole. Quantum dot
electron-hole exchange splits the ``dark'' and ``bright'' lines. Thus, in the
lowest energy $F=1$ manifold there are 6 energy lines in the XSM corresponding
to 3 projections of $F_z$ and the two excitons.

The GSM features now an ionized Mn acceptor interacting with a quantum dot hole
instead of with an acceptor hole. Both the strength and the symmetry of this
coupling are different: the QD hole-Mn coupling is much weaker than acceptor hole-Mn coupling and, due to the lack of spherical symmetry of the QD hole
state, is predominantly Ising \cite{Besombes04,JFR06}. Hence, the GSM is given
by the Ising coupling of the ionized Mn, with spin $S=5/2$ and the QD hole, with
total angular momentum $\sigma_h=\pm 3/2$. The spectrum of this system, relevant
for single Mn in CdTe QD \cite{JFR06,JFR07}, is formed by six doublets. We can
label the GSM states with the projections of the Mn spin and the QD hole spin
along the growth axis, $|M_z,\sigma_h\rangle$. Their eigenvalues are:
\begin{equation}
E_{M_z,\sigma_h}= +J^G_{hM} M_z \sigma_h. \label{Ising}
\end{equation}
where $J^G_{hM}$ is the exchange coupling between the QD hole and the ionized Mn spin.

There could be as many as 36 PL spectral lines joining the 6 energy levels of
the XSM and the 6 energy levels in the GSM. The highest energy PL would
correspond to the highest energy exciton state (within the $F=1$ manifold), with
quantum numbers $F_z=0$, $X=\pm 1$ and the lowest energy ground state, with
quantum numbers $M_z =+5/2,\sigma_h=-3/2$ or $M_z =-5/2,\sigma_h=+3/2$.
Interestingly, the $M_z=\pm 5/2$ state has zero overlap with the $F_z=0$, so
that this particular transition is forbidden (see figure
\ref{bandtoacceptor-scheme}). Going up in the $M_z$ ladder, the first excited
states in the GSM are $M_z=+3/2,\sigma_h=-1/2$ and $M_z =-3/2,\sigma_h=+1/2$. 

\begin{figure}[hbt]
\includegraphics[width=\columnwidth]{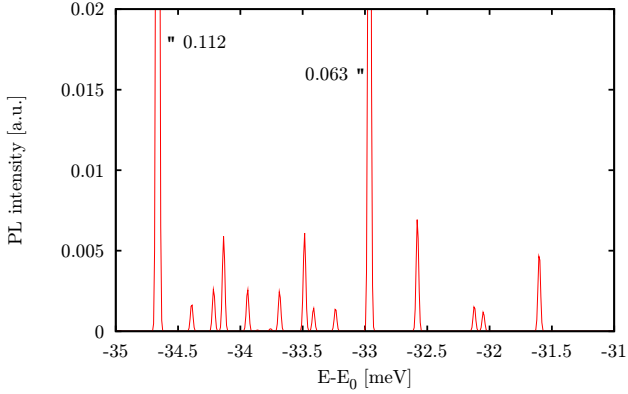}
\caption{$B_z=0$ $\sigma^+$-PL for neutral exciton band-to-acceptor transition, as calculated with the 4-spin-model. (Color online)}
\label{bandtoacceptor-PL-4}
\end{figure}

The results of the simulation within the 4-spin-model are shown in figure (\ref{bandtoacceptor-PL-4}). 
Using values from the band to band transition we take $D=4.7$ meV, $E=0.31$ meV,
 $\epsilon=5$ meV, $J_{hM}=-0.0405$ meV, $J_{eh}=-0.2$ meV. We assume that the QD hole-acceptor hole coupling is zero, but we take a ferromagnetic coupling between the Mn and the QD  hole to reproduce the effective ferromagnetic coupling between the exciton and the Mn acceptor complex. In the ground state we take a QD hole-Mn coupling larger than that of the XSM to account for the larger electrostatic attraction of the ionized Mn acceptor, $J^G_{hM}=0.15$ meV. 
The PL features a group of higher intesity lines at low energy and a group of weaker lines  2 meV above. As we discuss below, the splitting between the low and the high energy groups turns out to be given by ${\cal D}=3D/10$. Thus, band-to-acceptor transitions would permit a direct spectroscopic measurement of $D$. The V shape of the intensity pattern in the low energy group is somewhat related to the PL spectra of charged Mn-doped CdTe quantum dots \cite{Leger06}. In both cases the optical matrix elements feature overlap between states with well defined Mn spin and states where the Mn spin is Heisenberg exchanged coupled to another carrier (the extra electron in Mn doped CdTe and the acceptor hole in the case considered here).

\subsection{Band-to-acceptor transition: 2-spin-model}
The results of figure (\ref{bandtoacceptor-PL-4}) can be rationalized with the 2-spin-model,
 like in the band-to-band transition. We describe exciton states as the product of the QD electron-hole
pair spin, which can be $\pm 2$ or $\pm 1$ times the acceptor complex spin
$F_z=\pm 1, 0$. We ignore, for a moment, the in-plane anisotropy term ${\cal E}$
that mixes $F_z=\pm 1$. Thus $F_z$, $\sigma_c$ and $\sigma_h$ are good quantum
numbers in the XSM and $M_z$ and $\sigma_h$ are good quantum numbers in the GSM.
Of course, the spin of the QD hole, $\sigma_h$, and the Mn spin, $M$, are conserved during the
transition.

The quantum matrix elements of these transitions are given by the matrix elements of the dipole operator (eq. \ref{b2aplus} and \ref{b2aminus}) between emitting states, $X$ and ground
states with quantum numbers $M_z,\sigma_h$:
\begin{eqnarray}
|\langle M_z,\sigma_h|{\cal P}_{+}| \Psi_X\rangle|^2 =
 | {\cal C}^X_{M_z,+3/2,\downarrow ,\sigma_h} 
 + \frac{1}{\sqrt{3}}{\cal C}^X_{M_z,+1/2,\uparrow ,\sigma_h} |^2 
 \nonumber \\
|\langle M_z,\sigma_h|{\cal P}_{-}| \Psi_X\rangle|^2 =
| {\cal C}^X_{M_z,-3/2,\uparrow ,\sigma_h} +
 \frac{1}{\sqrt{3}}{\cal C}^X_{M_z,-1/2,\downarrow ,\sigma_h} |^2 \nonumber
\end{eqnarray}
for the $\sigma^+$ and $\sigma^-$ transitions respectively. 
Ignoring the in-plane mixing term ${\cal E}$, for a given QD exciton
($\sigma_c,\sigma_h$), the coefficients ${\cal C}^X_{M_z,j_h,\sigma_c,\sigma_h}$
are given by equation (\ref{F1}). Thus, as  long as $F_z$ is a good
 quantum number, the spin of the acceptor hole and the Mn satisfies the rule
\begin{equation}
M_z = F_z-j_z \label{rule2}
\end{equation}

Thus, for a given state in the XSM, with quantum numbers $|F=1,F_z,\sigma_c,\sigma_h\rangle$,
and a given polarization of the photon, $\pm 1$, we immediately get the
permitted values of the annihilated acceptor hole spin 
$j_z=\pm 1-\sigma_c$ (eq. \ref{rule1})
and the  allowed values of the Mn spin after photon emission (eq. \ref{rule2}).

Using equations (\ref{F1}), (\ref{rule1}),
(\ref{rule2}) one can make a table where the inputs are the QD electron spin
$\sigma_c$, the acceptor complex spin $F_z$, the circular polarization of the
photon $\sigma^{\pm}$, and the outputs are the annihilated acceptor hole spin $j_z$, the Mn spin $M_z$ after photon
recombination and the quantum yield of the transition,
${\cal I}\equiv I(F_z,\sigma_c,\sigma^{\pm},M_z)$. For $\sigma^+$ polarization we obtain:
\begin{eqnarray}
\begin{array}{r|r||r|r||l}
F_z & \sigma_c&  j_z& M_z & {\cal I}(\sigma^+) \\
\hline \hline
+1 & \uparrow & +1/2  &+1/2 & 5/100 \\
0  & \uparrow & +1/2  & -1/2 & 10/100 \\
-1 & \uparrow & +1/2  & -3/2 & 10/100 \\
+1 & \downarrow  &+3/2  & -1/2 & 5/100 \\
0 & \downarrow   &+3/2  & -3/2 & 20/100 \\
-1 & \downarrow &+3/2  & -5/2 &50/100 
\end{array}
\label{table}
\end{eqnarray}
The table for $\sigma^-$ is obtained by application of the time reversal operator, which changes $F_z, \sigma_c,j_z,M_z,$ to 
$-F_z, -\sigma_c,-j_z,-M_z$, and give the same intensity. 
In order to get the PL spectrum we need the transition energies $E_X-E_G$. 
Neglecting the ${\cal E}$ term that mixes $F_z=\pm 1$, the XSM  energies are approximated by
\begin{equation}
E_X(F_z,\sigma_c,\sigma_h) = E_0 -{\cal D}F_z^2 + {\cal J} X_z F_z + J_{eh} \sigma_c \sigma_h
\end{equation}
and the energy of the ground states by equation (\ref{Ising}). Thus, the transition energies $\omega=E_X-E_G$ are given by:
\begin{equation}
%E_{X\rightarrow G}=
\omega= E_0 -{\cal D}F_z^2 + 
{\cal J} (\sigma_c+\sigma_h) F_z + \sigma_h( J_{eh} \sigma_c 
-J^G_{hM}M_z)
\end{equation}

\begin{figure}[hbt]
\includegraphics[width=\columnwidth]{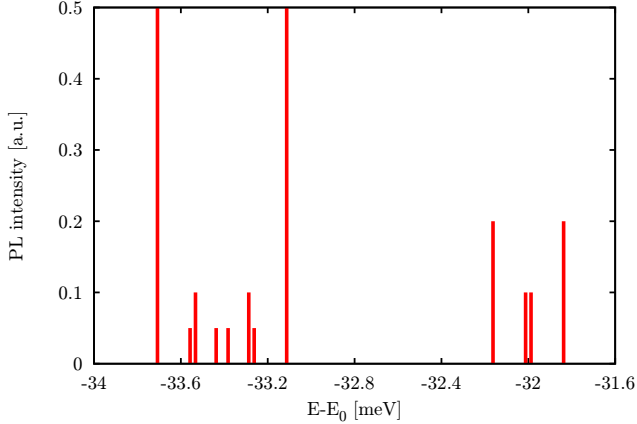}
\caption{$B_z=0$ $\sigma^+$-PL for neutral exciton band-to-acceptor transition, 
calculated with the  2-spin-model with ${\cal E}=0$ (see text). (Color online)}
\label{bandtoacceptor-PL}
\end{figure}

In figure (\ref{bandtoacceptor-PL}) we plot the intensity of the lines, without thermal depletion of the higher energy $F_z=0$ states.
The numerical values of the constants are those of the 4-spin-model simulation, properly renormalized to the 2-spin-model. Thus
we take ${\cal D}=3 D/10=1.41$ meV,
 ${\cal J}=\frac{21J_{hM}}{8}=-0.12$ meV. The numerical values of the QD electron-hole exchange and the QD hole-Mn coupling in the ground state are
the same in the 2-spin and 4-spin models,   $J_{eh}=-0.2$ meV and $J^G_{hM}=+0.15$meV. 
We take $J^G_{hM}>{\cal J}$ because the quantum dot hole is electrostatically attracted towards the Mn when this is ionized. The PL of figure (\ref{bandtoacceptor-PL}) features 12 peaks, corresponding to the 6 allowed transitions of equation (\ref{table}) times the two possible spin orientations of the QD hole. The high energy group corresponds to the transitions coming from the $F_z=0$ states. Most of the splitting between the high and the low energy group  is given by ${\cal D}$, with some contribution coming from the exciton Mn-complex exchange.   
 The lower energy group of 8 peaks correspond to transitions in the $F_z=\pm 1$ manifold. The high intensity peaks correspond to transitions where the Mn spin is left in a $\pm 5/2$ state. For $\sigma^+$ transitions, shown in the figure, the peaks correspond to final states with $M_z=-5/2$ and the quantum dot hole with spin $+1/2$ (low energy peak) and $-1/2$ (high energy peak). The splitting between those two is $\frac{5 J_{vM}}{2} +\frac{J_{eh}}{2} - {\cal J}$. Interestingly, this simple model captures the main features of the PL, as calculated with the 4-spin-model. The latter has more lines because of the the ${\cal E}$ terms, that mix $F_z=1$ to the $F_z=-1$ terms.

It is worth noting that the band-to-acceptor transitions permit to relate the
degree of spontaneous circular polarization to a possible inbalance in the $F_z$
population\cite{Awschalom08,Averkiev,Karlik}. For instance, the transitions that have has
initial state $F_z=-1$, averaged over all the possible QD spin orientation and
final state have a degree of cicular polarization of more than 70 percent:
\begin{equation}
\frac{\sum_{M_z,\sigma_c} \left(I(-1,\sigma_c,M_z,+)-I(-1,\sigma_c,M_z,-)\right)}
{\sum_{M_z,\sigma_c} \left(I(-1,\sigma_c,M_z,+)+I(-1,\sigma_c,M_z,-)\right)} =\frac{5}{7} 
\end{equation}
Thus the degree of Mn spin polarization can be probed by measuring the degree of circular polarization of the PL.

Hence, whereas band-to-band transitions probe the Mn-acceptor complex, which behaves as a spin $F=1$ object, the band-to-acceptor transitions connect initial states for which the Mn spin is correlated to the acceptor hole to final states where the Mn spin is Ising coupled to the QD hole, but with $M_z$ as a good quantum number. In band-to-acceptor transitions the photon energy and polarization carry information about $M_z$.

%%%%%%%%%%%%%%%%%%%%%%%%%%%%%%%%%%%%%%%%%%%%%%%%%%%%%%%%%%%%%%%%%%
\section{Charged exciton spectroscopy}
We now consider the PL spectra of negatively charged InAs dots. From the theory
side there is a lot of interest on the effect of number of carriers on the
magnetic properties of a dot doped with Mn atoms
\cite{Efros,JFR07,JFR04,Pawel-1,Climente05,Abolfath07,Peeters07}. A priori, the
ground state of a negatively charged InAs dot doped with a single Mn should be
the ionized $A^-$ acceptor, with spin properties identical to those of Mn in
neutral CdTe \cite{JFR06}. Band-to-band transitions should yield zero field PL
spectra with 6 peaks. Band-to-acceptor transitions for the negatively charged
dot, $e^-A^0 \rightarrow A^-$, have been studied by by A. Govorov in reference
\onlinecite{Govorov04}. He obtained a PL spectrum with 3 peaks in at zero
magnetic field.

\begin{figure}[hbt]
\includegraphics[width=0.8\columnwidth]{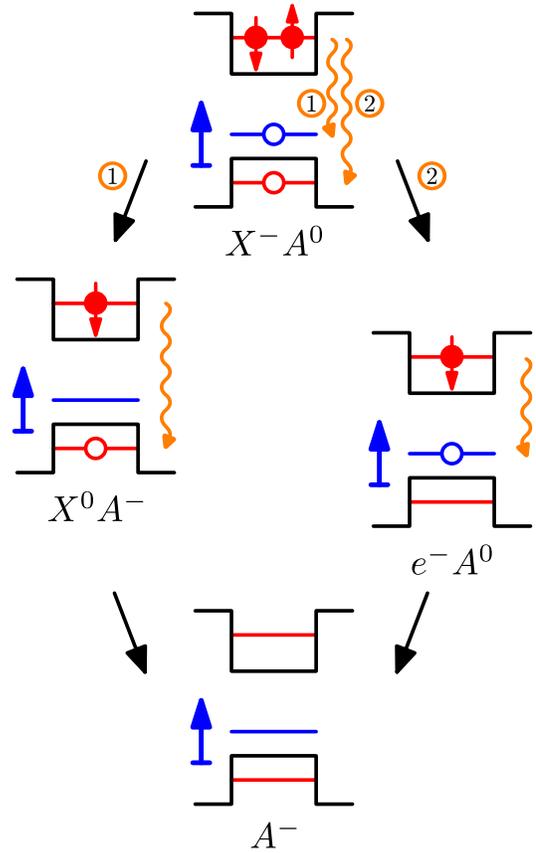}
\caption{Possible electronic configurations for two photon decay of the negatively charged InAs QD with one Mn. The uppermost diagram shows the configuration with two QD electrons, one QD hole and the neutral Mn-acceptor hole complex. By a band-to-acceptor transition (pathway 1), one arrives at one of the intermediate configurations, with an ionized Mn and a QD exciton. A further band-to-band transition, creates the lowest energy configuration: an ionized Mn. Another recombination possibility is via pathway 2: a band-to-band transition yields the intermediate configuration with the neutral Mn-acceptor complex and one QD electron. A further band-to-acceptor transition then yields the same end configuration as the one achieved via pathway 1. (Color online)}
\label{4scheme}
\end{figure}

Contrary to these expectations, the experimental results on negatively charged
single Mn doped InAs QD \cite{Voisin07} show very similar neutral and charged
exciton band-to-band spectra, with 5 peaks in the zero field PL. The reported
transitions correspond to an emitting state with a neutral Mn acceptor plus a
quantum dot trion ($X^-A^0$). Thus, there are two electrons in the conduction
level, and two holes, the acceptor hole and the QD hole (see figure
\ref{4scheme}). The final state, after the emission of two photons, is the
ionized Mn, with spin $S=5/2$. The fact that the reported neutral and negatively
charged transitions are very similar indicates that QD electrons are very weakly
coupled to the Mn-acceptor hole complex. This is different from CdTe dots doped with
one Mn, where the addition of a single electron changes the spin properties of
the Mn \cite{JFR04,Leger06,JFR07}. In InAs, the Mn is strongly coupled to the
acceptor hole and is less sensitive to the number of conduction band carriers.

Yet the presence of additional QD electrons can result in new PL spectra when we
consider band-to-acceptor transitions, unreported so far in single Mn-doped InAs
quantum dots. Since the negatively charged trion $X^-A^0$ features 2 electrons
and 2 holes in different states, the decay towards the ground state $A^-$ can
occur via two recombination pathways, as show in figure \ref{4scheme}. One of
the pathways (denoted as 1 in the figure) consist in a band-to-acceptor
transition $X^-A^0\rightarrow X^0A^-$ followed by a $X^0A^-\rightarrow A^-$,
i.e., a band-to-band transition that annihilates a QD exciton coupled to a
ionized Mn acceptor. The PL of the first step in pathway 1 is related to the
band-to-acceptor transition of the neutral dot discussed in the previous
section. The PL of the second step in pathway 1 should be very similar to that
of single Mn doped CdTe QD. As long as the Ising part of the coupling between
the QD hole and the Mn spin is dominant, as in CdTe, the PL of the energy and
polarization of the photon yield direct information of the Mn spin after the
exciton recombination \cite{JFR06}.

Pathway 2 in the figure starts with a band-to-band transition $X^-A^0\rightarrow
e^-A^0$ followed by a band-to-acceptor transition $e^-A^0 \rightarrow A^-$. The
first step has been observed experimentally and, since the electron-Mn exchange
is very weak, yields a PL very similar to the neutral case considered above. The
second step is identical to the transition considered by Govorov
\cite{Govorov04}.

\begin{figure}[hbt]
\includegraphics[width=\columnwidth]{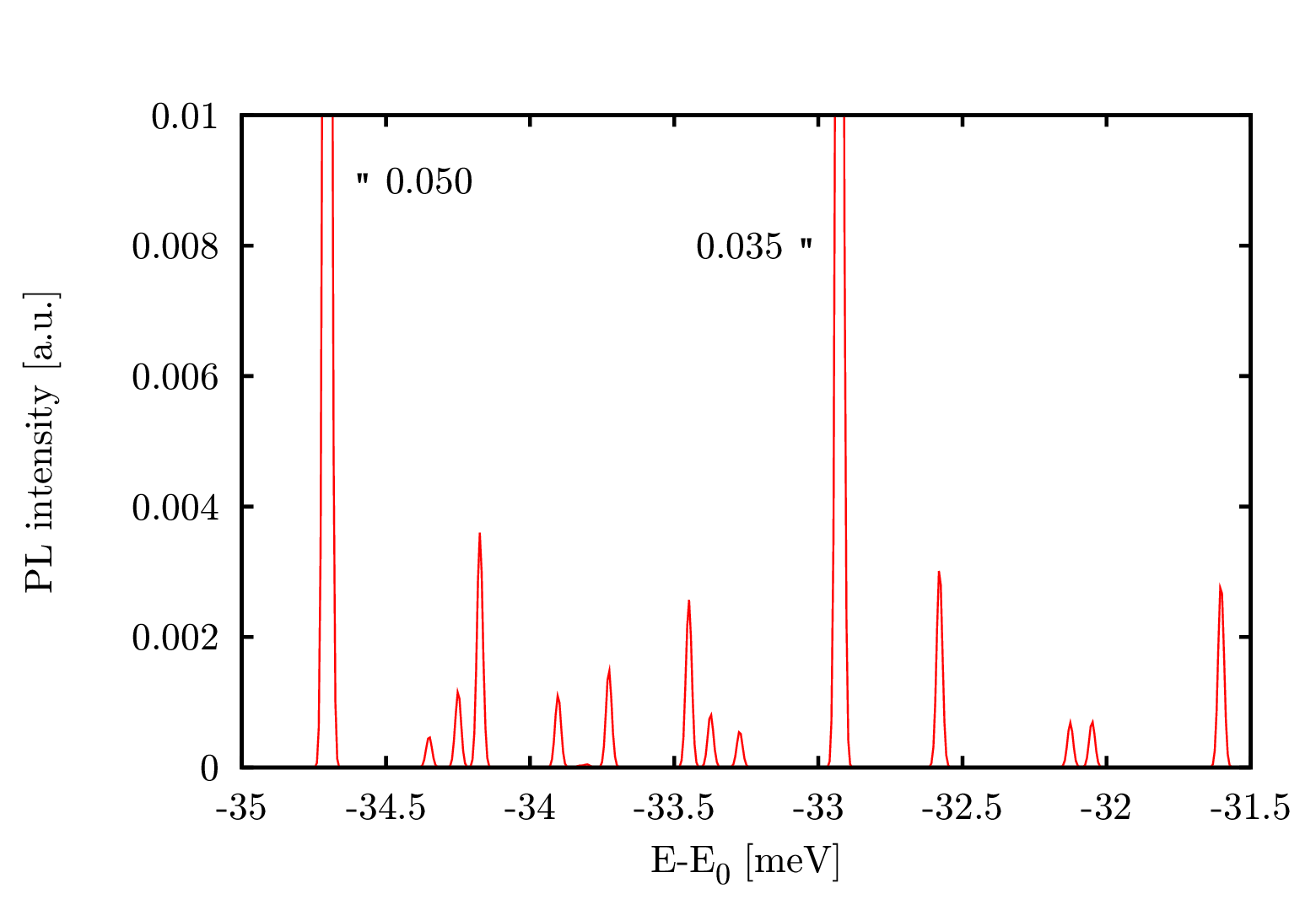}
\includegraphics[width=\columnwidth]{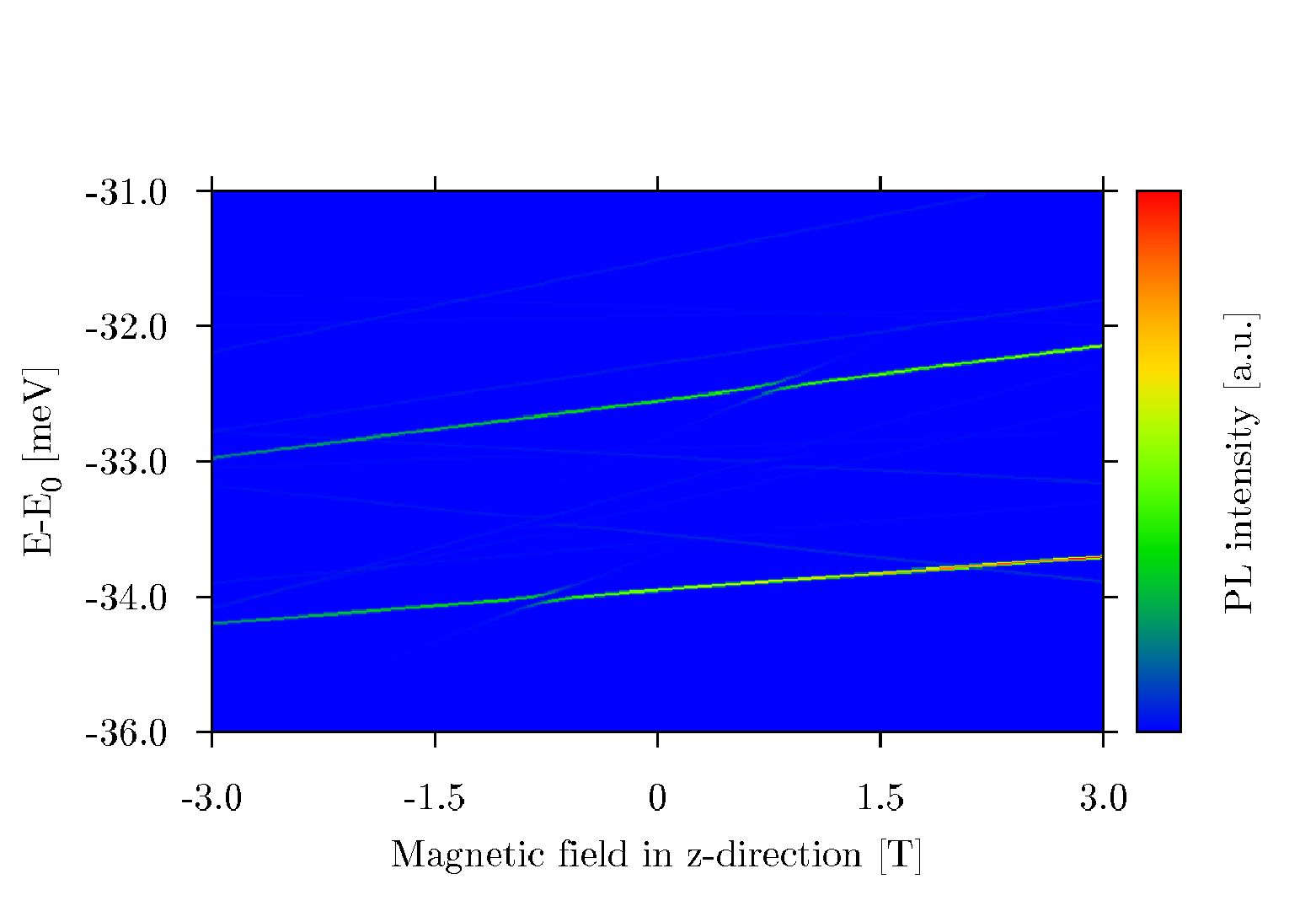}
\caption{Upper panel: $B_z=0$ $\sigma^+$-PL for a negatively charged exciton band-to-acceptor transition ($X^-A^0 \rightarrow X^0A^-$), as calculated with the 4-spin-model. Lower panel: Intensity plot for the $\sigma^+$-PL for the negatively charged exciton band-to-acceptor transition ($X^-A^0 \rightarrow X^0A^-$), as a function of energy (vertical axis) and applied magnetic field (horizontal axis). (Color online)}
\label{Paul1}
\end{figure}

In the upper panel of figure (\ref{Paul1}) we show the zero field PL for the
band-to-acceptor transition for negatively charged InAs QD doped with 1 Mn. The
emitting state is $X^-A^0$ and the final state is $X^0A^-$, a quantum dot
exciton coupled to an ionized Mn acceptor. The corresponding energy level
diagram is very similar to that of the neutral band-to acceptor transition 
shown in figure
(\ref{bandtoacceptor-scheme}).  Ignoring states with $F>1$, there
are 6 exciton states, corresponding to 3 $F_z$ values and 2 QD hole spin states.
There are 24 ground states, corresponding to the 6 Mn spin orientations and the 4
spin states of the quantum dot exciton. This is in contrast with the $X^0A^0
\rightarrow h^+A^-$ transition, for which both the GSM and the XSM have 12
states. Other differences with that transition is the lack of QD electron-hole
exchange in the emitting state, and the presence of that coupling in the ground
state. Thus, electron-hole exchange is present both in the neutral and in the
charged case, either in the XSM or in the GSM. It would be possible to do an
analytical model for the charged case along the lines of the previous section
for which the optical matrix elements would be still given by equation
(\ref{table}). Not surprisingly, the PL $X^-A^0\rightarrow X^0A^-$ of figure
(\ref{Paul1}), calculated with the 4-spin-model, is quite similar to that of the
$X^0A^0 \rightarrow h^+A^-$.

In the lower panel of figure (\ref{Paul1}) we plot the $\sigma^+$ PL as a function of the
magnetic field in the Faraday configuration. The two low energy brighter peaks
correspond to transitions where the Mn spin is, in the final state $M_z=-5/2$.
They are splitted due to the different spin orientations of the QD exciton to
which they are coupled. This results also in different slopes as the magnetic
field is ramped.  The evolution of the energy levels in
the ground state results in a compensation of the zero field exchange splittings
by the finite field Zeeman splittings. At a particular value of the magnetic
field, several lines become degenerate. In the absence of spin-flip terms, they
do not anti-cross. This particular energy arrangement has been reported in CdTe
doped with Mn \cite{Besombes04}. The anti-crossings observed at $B_z \simeq 1$T
are related to those taking place at the XSM and dicussed above in the context
of neutral band-to-band transitions.

%\begin{figure}[hbt]
%\includegraphics[width=\columnwidth]{X-A0_X0A-_plus_B_linear_low_res}
%\caption{
%Color plot for the circularly polarized $\sigma^+$ PL for the negatively charged
%exciton band-to-acceptor transition ($X^-A^0 \rightarrow X^0A^-$), as a function
%of energy (vertical axis) and applied magnetic field (horizontal axis). (Color
%online)}
 % Lower panel: same than upper panel, in logarithmic scale appreciate weaker intensiy peaks.
%\label{Paul-field}
%\end{figure}

%\onecolumngrid
%\begin{widetext}
%\begin{center}
%\begin{figure}[hbt]
%\includegraphics[width=0.8\textwidth]{scheme_X-A0_X0A-}
%\caption{Energy level diagram for the charged exciton band-to-acceptor
%transition ($X^-A^0 \rightarrow X^0A^-$). Both the ground state and exiton state
%energy lines are shown, as well as their evolution as a function of the magnetic
%field. (Color online)}
%\label{Paulscheme}
%\end{figure}
%\end{center}
%\end{widetext}
\twocolumngrid

%\begin{figure}[h]
%\begin{tabular}{cc}
% \includegraphics[angle=-90,width=0.45\textwidth]{X-A0_X0A-_plus_B=0_linear_zoom} &
% \includegraphics[angle=-90,width=0.45\textwidth]{X-A0_X0A-_linear_B=0_linear} \\
% \includegraphics[angle=-90,width=0.45\textwidth]{X-A0_X0A-_plus_B_log} & 
% \includegraphics[angle=-90,width=0.45\textwidth]{X-A0_X0A-_linear_B_log} \\
% \multicolumn{2}{c}{\includegraphics[angle=-90,width=0.9\textwidth]{scheme_X-A0_X0A-}} \\
%\end{tabular}
%\end{figure}

\section{Discussion and conclusions}
We have addressed the problem of single exciton spectroscopy of a single-Mn InAs
doped quantum dot. The main goal is to link a few-spin Hamiltonian with the PL
spectra featuring spin-split peaks at zero magnetic field. We have focused on
the fact that, for single Mn-doped InAs QD this is a 4 body problem, with the QD
electron, QD hole, acceptor hole and Mn spin. 
In order to account for the experimental observations \cite{Voisin07} it is 
important to assume that a bulk-like acceptor states surives inside the gap, 
weakly affected by the quantum dot.
The strongest exchange interaction
is that of the Mn and the acceptor hole. In most instances this permits to
interpret the results as if the quantum dot exciton interacts with a spin $F=1$
object, obtained from the antiferromagnetic coupling of the Mn spin $S=5/2$ and
the acceptor hole $j=3/2$. We use both a 4-spin-model, in which the identity of
the spin of all the carriers is included in the calculation, and a 2-spin-model,
that ignores the composite nature of the exciton and the Mn-acceptor complex.
The models are diagonalized numerically and the PL spectra are obtained, taking
full account of the optical and spin selection rules.

The 2-spin-model portraits the Mn acceptor complex in InAs as a spin $F=1$
nanomagnet with two almost degenarate ground states, $F_z=\pm 1$, a rather large
single ion magnetic anisotropy of ${\cal D}$, and a small in-plane anisotropy
${\cal E}$.  The notion that single Mn atoms in
quantum dots can behave like aritifical single molecule magnets has been
discussed before \cite{chamonix,JFR07}. Interestingly, the zero field
exciton spectroscopy gives a direct measurement of
${\cal E} \simeq 0.035$ meV, but only 
an indirect measurement of $D \simeq 4$ meV,
through the height of the central peak.  The evolution of the PL spectra as a
magnetic field is applied permits an indirect measurement of other energy scales
in the problem. In the band-to-band transitions, measured by Kudelski et al.
\cite{Voisin07}, there is particular value of the field $B^*$ at which three
lines in the spectra merge. $B^*$ is the field at which
 the Zeeman splitting and exchange coupling to the exciton have the same
 intensity and opposite sign (eq. \ref{Bstar}).
Since the zero field measurement provides ${\cal  J}$, the value of
 $g_F$  in InAs can be inferred from $B^*$. We estimate
$g_F=3.01$. 
  
The nature of Mn spin $S=5/2$ can be unveiled in two manners: upon electron
doping the system and in band-to-acceptor recombination of the charge neutral
dot. The latter results in the ionization of the Mn complex so that, in the
final state, the Mn spin is $S=5/2$ and is presumably Ising coupled to the
QD hole. The band-to-band and band-to-acceptor transitions are very
different. In the former the Mn spin is slaved by the acceptor hole both in the
XSM and GSM states. Thus, photon emission does not change the symmetry of the Mn
spin Hamiltonian. In the band-to-acceptor transition the Mn is liberated from
the acceptor hole in the final state, so that photon emission involves a change
of the effective Mn spin Hamiltonian. In this sense, the band-to-acceptor
transition resembles the negatively charged trion transition in Mn doped CdTe
\cite{Leger06}. Band-to-acceptor transitions would provide a direct
spectroscopic measurement of ${\cal D}$.

Notice that, within the 2-spin-model we take an effective ferromagnetic coupling
between the exciton and the Mn. The 2-spin-model does not say whether this
coupling is QD hole-Mn acceptor complex, QD hole-acceptor hole, QD electron-Mn or QD electron-acceptor
hole. If we assume that the dominant coupling is between the Mn spin and the QD
hole, we could conclude that the coupling is ferromagnetic, at odds with the
usual antiferromagnetic coupling of holes and Mn in III-V materials.
Interestingly, we have seen how the bare sign of the QD hole-Mn coupling is
reversed when going from the 4-spin-model to the 2-spin-model (see equation
\ref{sign-reverse}). The strong Mn-acceptor hole complex results in the
renormalization of the spin interactions of these two spins with applied field
and the exciton spins. This can even result in sign inversion of the exchange
interaction. Thus, we have shown that a ferromagnetic coupling between the 2
spins of the simpler model can arise even if the underlying spin couplings of
the QD hole to the Mn spin are antiferromagnetic. In this way, we reconcile
the standard view about this system, in which the holes are coupled
antiferromagnetically to the Mn spin, with the the observations \cite{Voisin07}.

A single Mn in a quantum dot provides an ideal system to address and control the
spin of a single object in a solid state environement. The PL spectra provide
valuable information of the effective spin Hamiltonians, for different states of
the dot. A Mn atom in a InAs QD behaves like a 3 level system. It might be
possible to use the encode a single qubit in the almost degenerate $F_z=\pm 1$
doublet. The presence of higher energy $F_z=0$ state, acting as a barrier, might
block spin relaxation between the ground states. In analogy with the case of
electron doped and hole doped quantum dots, it should be possible to manipulate
the spin of a single or a few Mn spins in a quantum dot by application of laser
pulses.

We acknowledge fruitful discussions with A. Govorov.
This work has been financially supported by MEC-Spain (Grants
FIS200402356 and the Ramon y Cajal Program) and  
by Consolider CSD2007-0010 and, in part, by FEDER funds.

\appendix
\section{Analytical solution of the 2-spin-model}
Here we provide an analytical solution of the ground state and exciton state manifolds within the 2-spin-model. The GSM has 3 states, but the Hamiltonian can be block diagonalized since the $F_z=0$ state is decoupled from the $F_z=\pm 1$ doublet. The XSM state has 12 states, corresponding to the 4 spin orientations of the exciton and the 3 states of $F_z$ in the $F=1$ manifold. Since we consider Ising interaction between the exciton and the $F$ spin, the Hamiltonian of the XSM is also block diagonalized. Importantly, both the GSM and the XSM are each characterized by a single angle, $\theta_G$ and $\theta_X$, that characterize the ratio between the in-plane mixing of the $F_z=\pm 1$ components and their splitting, induced both by exchange interaction with the exciton and by Zeeman coupling. As we show here, the lineshape of the PL spectrum depends on $\frac{1}{2} \left( \theta_X-\theta_G \right)$.

In the basis $|1,+1\rangle,|1,-1\rangle,|1,0\rangle$ the Hamiltonian of the GSM reads:
\begin{eqnarray}
{\cal H}_G = \left(\begin{array}{ccc} -{\cal D} + g_F\mu_B B_z & {\cal E} & 0 \\
 {\cal E} & -{\cal D}-g_F\mu_B B_z & 0 \\
 0 & 0 & 0 \end{array} \right) 
\label{3by3}
\end{eqnarray}
We can write the two by two matrix within the $F_z=\pm 1$ subspace as a linear combination of the unit matrix $\bm I$, and the Pauli matrices $\bm \sigma_z$ and $\bm \sigma_x$: 
\begin{equation}
{\cal H}_{G, \pm 1}= -{\cal D} \bm I + \vec h_{G} \left(\cos(\theta_G) \bm \sigma_z + \sin(\theta_G) \bm \sigma_x\right) 
\end{equation}
with
\begin{equation}
{\vec h}_G= ({\cal E},g_F\mu_B B_z)=h_G \left(\sin(\theta_G),\cos(\theta_G)\right)
\end{equation}
where $h_G=\sqrt{{\cal E}^2 +(g\mu_B B_z )^2}$. Thus, we have 
\begin{equation}
\cot(\theta_G)=\frac{g_F\mu_B B_z}{{\cal E}} 
\end{equation}
At zero field we have $\theta_G=\pi/2$. 

The eigenvalues of the GSM are: $-{\cal D}-h_G$, $-{\cal D}+h_G$ and $0$. The corresponding eigenvectors are the product the quantum dot ground state, denoted by $|0\rangle$, and the spin part, denoted by $\Psi_G^-$, $\Psi_G^+$ and $\Psi_G^0$ respectively. The spin part of the $F_z=\pm 1$ sector reads
\begin{eqnarray}
\left(\begin{array}{c} \Psi_{G}^- \\ \Psi_{G}^+ \end{array}\right)= 
\left(\begin{array}{cc}
\sin\left(\frac{\theta_G}{2}\right) & -\cos\left(\frac{\theta_G}{2}\right) \\
\cos\left(\frac{\theta_G}{2}\right) & \sin\left(\frac{\theta_G}{2}\right) \\
  \end{array}\right)
\left(\begin{array}{c} |1,+1\rangle \\ |1,-1\rangle \end{array}\right) 
\end{eqnarray}
Notice that, in the limit of very strong field, $g_F\mu_B B_z \gg {\cal E}$, the mixing between $F_z=+1$ and $-1$ vanishes.

We now consider the XSM, which is split in 4 sectors with 3 states each and a well defined exciton state $X_z=\pm 1$ and $X_z=\pm 2$. We focus on the optically active excitons, $X_z=\pm 1$. Since $X_z$ commutes with the XSM Hamiltonian, the $X_z=-1$ and $X_z=+1$ sectors decouple and are described by 3 by 3 matrices with the same structure as (\ref{3by3}). The $F_z=0 \otimes X_z$ states are decoupled from the $F_z=\pm 1 \otimes X_z$ states. In the basis $|1,+1\rangle\otimes |X_z\rangle$, $|1,-1\rangle\otimes |X_z\rangle$ the XSM Hamiltonian for exciton $X_z$ reads:
\begin{eqnarray}
{\cal H}_{X_z=\pm 1} &=& \left(E(X_z)-{\cal D}\right) \bm I \nonumber \\
&&+ {\vec h}_X \left(\cos(\theta_X) \bm \sigma_z + \sin(\theta_X) \bm \sigma_x\right) 
\end{eqnarray}
where $E_X(X_z)= E_0 + g_X \mu_B X_z B_z$ is the energy of the excitonic transition neglecting spin coulings plus the QD exciton Zeeman splitting, and
\begin{eqnarray}
{\vec h}_X &=& ({\cal E},g_F\mu_B B_z + {\cal J} X_z) \nonumber \\
&=& h_X \left(\sin(\theta_X),\cos(\theta_X)\right)
\end{eqnarray}
where
\begin{equation}
h_X=\sqrt{{\cal E}^2 + \left(g_F \mu_B B_z + {\cal J} X_z \right)^2} 
\end{equation}
Now we have
\begin{equation}
\cot(\theta_X)=\frac{g_F \mu_B B_z + {\cal J} X_z}{{\cal E}}
\end{equation}
Notice that the angle $\theta_X$ depends both on the magnetic field $B_z$ and on the exciton spin projection, $X_z=\pm 1$. At zero field the angles corresponding to $X_z=+1$ and $X_z=-1$ differ by $\pi$, so that $\cot(\theta_X) = \pm \frac{{\cal J}}{{\cal E}}$. At finite fields the effect of exchange and magnetic field on the Mn-acceptor hole complex can either compete or cooperate and the relation between $\theta_X$ for $X_z=\pm 1$ is non-trivial.

The eigenvalues are for the XSM are:
\begin{eqnarray}
&&E_0 + g_X \mu_B X_z B_z - {\cal D} - h_X \nonumber \\
&&E_0 + g_X \mu_B X_z B_z - {\cal D} + h_X \\
&&E_0 + g_X \mu_B X_z B_z \nonumber
\end{eqnarray}
The corresponding eigenvectors are the product of the quantum dot exciton, denoted by $|X_Z\rangle$ and the spin part, denoted by $\Psi_X^-$, $\Psi_X^+$, $\Psi_X^0$ respectively. The spin part of the XSM eigenvectors in the $F_z=\pm 1$ sector are:
\begin{eqnarray}
\left(\begin{array}{c} \Psi_X^- \\ \Psi_X^+ \end{array}\right)= 
\left(\begin{array}{cc}
\sin\left(\frac{\theta_X}{2}\right) & -\cos\left(\frac{\theta_X}{2}\right) \\
\cos\left(\frac{\theta_X}{2}\right) & \sin\left(\frac{\theta_X}{2}\right) \\
  \end{array}\right)
\left(\begin{array}{c} |1,+1\rangle \\ |1,-1\rangle \end{array}\right)
\end{eqnarray}

The intensity of the PL lines is given by:
\begin{equation}
 I_{\pm}(\omega)= \sum_{a,b} n_b |\langle \Psi_{G}^a |\Psi_{X}^b\rangle|^2
\delta(E_X^b-E_G^a +\omega)
\label{A12}
\end{equation}
where both $a$ and $b$ run over $+1,-1,0$. Here $n_b$ is the statistical occupation of the exciton states. Both the matrix elements and the allowed transitions depend on the exciton spin $X_z=\pm 1$, which is in turn given by the photon polarization. The intensity table $|\langle \Psi_{G}^a |\Psi_{X}^b\rangle|^2$ reads:
\begin{eqnarray}
\begin{array}{l||l|l|l}
_{\hspace{-0.cm}\Psi_G}^{\hspace{+0.3cm}\Psi_X}  & - & + & 0 \\
\hline \hline
- & \cos^2\left(\frac{1}{2}\left(\theta_G-\theta_X\right)\right) & \sin^2\left(\frac{1}{2}\left(\theta_G-\theta_X\right)\right) & 0 \\
+ & \sin^2\left(\frac{1}{2}\left(\theta_G-\theta_X\right)\right) & \cos^2\left(\frac{1}{2}\left(\theta_G-\theta_X\right)\right) & 0 \\
0 & 0 & 0 & 1 
\end{array} \nonumber \\ \nonumber
\label{table-PL-I}
\end{eqnarray}
This matrix has 5 non-zero elements, corresponding to the 5 permited transitions for a given circular polarization. The transition energies within the $F_z=\pm 1$ doublets are given by
\begin{equation}
E^{b\rightarrow a}= E_0+ g_X \mu_B X_z B_z + b h_X-a h_G
\label{2spin-energies} 
\end{equation}
where both $a$ and $b$ can take values equal to $\pm 1$.

As expected, the $0$ state is decoupled from the others and the optical matrix
element is 1. The 4 transitions on the $F_z=\pm 1$ doublet ($--,-+,+-,++$) are
governed by a single variable $\frac{1}{2}\left(\theta_G-\theta_X\right)$. The
intensity of the crossed sign transitions goes to zero in the limit of very
large fields or exchange coupling. In this situation the mixing induced by
in-plane anisotropy term, $E$, is negligible. Notice that the height of the
actual transitions depends both on $|\langle \Psi_{G}^a |\Psi_{X}^b\rangle|^2$
and on the statistical occupation. Thus, the observed intentisty of the $F_z=0$
line is smaller, due to a smaller statistical occupation.

%%%%%%%%%%%%%%%%%%%%%%%%%%%%%%%%%%%%%%%%%%%%%%%%%%%%%%%%%%%%%%%%%%

%\widetext

%\begin{figure}
%[t]
%\includegraphics[width=2.8in]{figure2b.eps}
%\caption{ \label{fig2}(Color online).
%(a) PL for dots $\#3$ (upper) and $\#4$ (middle and bottom). Influence of the
%LL-HH mixing and the $eh$ exchange on the PL. 
%(b) Zero field PL spectrum for dot $\#2$ with 3 Mn impurities
%as a function of the relative Mn-carrier coupling.}
%\end{figure}

%\begin{figure}
%[hbt]
%\includegraphics[width=2.8in]{figure2c.eps}
%\caption{ \label{fig2}(Color online).
% Upper panels: Influence of the
%LL-HH mixing and the $eh$ exchange on the PL of single Mn PL for dots
% $\#3$ (2a) and $\#4$ (2b and 2c). 
%2d-2g Zero field PL spectrum for dot $\#2$ with 3 Mn impurities
%as a function of the relative Mn-carrier couplings.}
%\end{figure}

%\begin{figure}
%[t]
%\centering
%\includegraphics[width=2.8in]{figure3.eps}
%\caption{ \label{fig3}(Color online).
%(a) Normalized histogram of Mn-hole coupling. The dashed line is the integrated
%histogram. (b) PL spectrum for 4 Mn impurities (see text) for several
%values of $B^z$. (c) Idem for 3 Mn impurities . (d)
%FWHM for as a function of $B^z$ and $T$=3,2,1 and 0.5 meV 
%(top to bottom).}
%\end{figure}

\end{document}